\documentclass[a4paper,fleqn,usenatbib]{mnras}
\usepackage{times}
\usepackage{mathptmx}

\usepackage{amsmath}    % Advanced maths commands
\usepackage{amssymb}    % Extra maths symbols

\usepackage{epsf,graphicx,rotating}
\usepackage{color}
\usepackage{hyperref}
\usepackage{xspace}

\newcommand{\mobh}{M_{\rm BH}}
\newcommand{\ledd}{\lambda_{\rm Edd}}
\newcommand{\lxray}{L_{\rm 2-10\,keV}}
\newcommand{\lbol}{L_{\rm bol}}
\newcommand{\ldisc}{L_{\rm disc}}
\newcommand{\bolc}{\kappa_{\rm 2-10\,keV}}

\newcommand{\xmmcosmos}{{\it XMM}-COSMOS\xspace}
\newcommand{\xmm}{{\it XMM-Newton}\xspace} 
\newcommand{\xxl}{XMM-XXL\xspace} 
\newcommand{\xxln}{XXL-N\xspace} 
\newcommand{\cha}{{\it Chandra}\xspace} 
\newcommand{\nh}{{$N_{\rm H}$}\xspace}

\title[X-ray properties of AGN in Northern XMM-XXL Field]{X-ray spectral properties of the AGN sample in the northern XMM-XXL field}
\author[Liu et al.,]{
Zhu Liu,$^{1,2,3}$\thanks{liuzhu@nao.cas.cn(ZL)}
Andrea Merloni,$^{3}$\thanks{am@mpe.mpg.de(AM)}
Antonis Georgakakis,$^{3,4}$
Marie-Luise Menzel,$^{3}$
\newauthor
Johannes Buchner,$^{3}$
Kirpal Nandra,$^{3}$ 
Mara Salvato,$^{3}$
Yue Shen,$^{5,6}$\thanks{Hubble Fellow}
Marcella Brusa,$^{3,7,8}$
\newauthor
Alina Streblyanska$^{9,10}$
\\
$^1$National Astronomical Observatories, Chinese Academy of Sciences, Beijing 100012, People's Republic of China\\
$^2$University of Chinese Academy of Sciences, Beijing 100049, People's Republic of China\\
$^3$Max-Planck-Institut f{\"u}r extraterrestrische Physik, Giessenbachstrasse 1, D-85748, Garching bei M{\"u}nchen, Germany\\
$^4$IAASARS, National Observatory of Athens, GR-15236 Penteli, Greece\\
$^5$Carnegie Observatories, 813 Santa Barbara Street, Pasadena, CA 91101, USA\\
$^6$Kavli Institute for Astronomy and Astrophysics, Peking University, Beijing 100871, People's Republic of China\\
$^7$Dipartimento di Fisica e Astronomia, Universit{\`a} di Bologna, viale Berti Pichat 6/2, I-40127 Bologna, Italy\\
$^8$INAF–Osservatorio Astronomico di Bologna, via Ranzani 1, I-40127 Bologna, Italy\\
$^9$Instituto de Astrofisica de Canarias (IAC), E-38200 La Laguna, Tenerife, Spain\\
$^{10}$Departamento de Astrofisica, Universidad de La Laguna (ULL), E-38205 La Laguna, Tenerife, Spain\\}

\date{Accepted XXX. Received YYY; in original form ZZZ}

\pubyear{2015}

\begin{document}
\label{firstpage}
\pagerange{\pageref{firstpage}--\pageref{lastpage}}
\maketitle

%%*************************************************
%%**************Section: Abstraction***************
%%*************************************************
\begin{abstract}

In this paper we describe and publicly release a catalogue consisting of 8445
point-like X-ray sources detected in the XMM-XXL north survey. For the 2512 AGN
which have reliable spectroscopy from SDSS-III/BOSS, we present the X-ray
spectral fitting which has been computed with a Bayesian approach. We have also
applied an X-ray spectral stacking method to different sub-samples, selected on
the basis of the AGN physical properties ($\lxray$, $z$, $\mobh$, $\ledd$ and
\nh). We confirm the well-known Iwasawa-Taniguchi effect in our
luminosity-redshift sub-samples, and argue that such an effect is due to a
decrease in the covering factor of a distant obscuring `torus' with
increasing X-ray luminosity. By comparing the distribution of the reflection
fraction, the ratio of the normalization of the reflected component to the
direct radiation, we find that the low-luminosity,
low-redshift sub-sample had systematically higher reflection fraction values than the
high-redshift, high-luminosity one. On the other hand, no significant
difference is found between samples having similar luminosity but different
redshift, suggesting that the structure of the torus does not evolve strongly
with redshift. Contrary to previous works, we do not find evidence for an
increasing photon index at high Eddington ratio. This may be an indication that
the structure of the accretion disc changes as the Eddington ratio
approaches unity. Comparing our X-ray spectral analysis results with the
optical spectral classification, we find that $\sim20$ per cent of optical type-1
AGN show an X-ray absorbing column density higher than $10^{21.5}\,\rm{cm^{-2}}$, and about 50
per cent of type-2 AGN have an X-ray absorbing column density less than
$10^{21.5}\,\rm{cm^{-2}}$. We suggest that the excess X-ray absorption shown
in the high-luminosity optical type-1 AGN can be due to small-scale dust-free
gas within (or close to) the broad line region, while in the low-luminosity
ones it can be due to a clumpy torus with a large covering factor.

\end{abstract}


%%*************************************************
%%**************Section: Introduction**************
%%*************************************************
\section{Introduction}

Active galactic nuclei (AGN) are among the most luminous objects in the
Universe and emit radiation over the whole range of the electromagnetic
spectrum.  They are believed to be powered by accretion of matter on to a super-massive black hole (SMBH). It is now well established that almost all massive
galaxies harbour a SMBH at their centre \citep{Kormendy+Richstone+1995}.  This
leads to the suggestion that most galaxies have been through a phase of nuclear
activity in the past \citep{Marconi+al+2004, Merloni+Heinz+2008, Shankar+2009},
and that SMBH growth is mainly due to accretion of matter during active
phases over cosmological times.

AGN are powerful X-ray emitters which dominate the source number counts in the
distant X-ray source populations. Because X-ray emission is less diluted
by the host-galaxy starlight and absorption of the material along the line of
sight, X-ray surveys of AGN provide the most complete and unbiased AGN census
\citep{Brandt+Alexander+2015}. Thanks to the improved sensitivity and
resolution of X-ray missions such as {\it X-ray Multi-Mirror Mission}
\citep[{\xmm}; e.g.][]{Jansen+al+2001} and {\it Chandra X-ray Observatory}
\citep[{\cha}; e.g.][]{Weisskopf+al+2000}, X-ray surveys are now able to detect
large numbers of AGN over a wide range of redshifts, providing a powerful tool
for studying the evolution of SMBH and their host galaxies. 

Observationally, the main feature in the X-ray spectra of AGN is a power-law
continuum, which is presented as $N(E)\propto E^{-\Gamma}$. This power-law
component is thought to be produced by inverse Comptonization of optical/UV
photons in a hot corona located close to the accretion disc
\citep[e.g.][]{Haardt+Maraschi+1991}. Investigations of the relation between
the power-law photon index, $\Gamma$, and the other AGN physical parameters,
e.g. the Eddington ratio $\ledd$\footnote{${\ledd} = L_{\rm bol}/L_{\rm Edd}$,
where $L_{\rm Edd} = 1.3\times10^{38}(\mobh/M_{\odot})$ and $L_{\rm bol}$ is
the bolometric luminosity.} and X-ray luminosity $L_{\rm X}$, may yield
important insights into the accretion processes and the corona properties.
Previous works have shown that $\Gamma$ is anti-correlated with $\ledd$ when
$\ledd\lesssim10^{-3}-10^{-2}$ \citep{Gu+Cao+2009} and strongly correlated with
$\ledd$ when $\ledd>10^{-3}-10^{-2}$ \citep{Lu+Yu+1999, Shemmer+al+2006,
Shemmer+al+2008, Wu+Gu+2008, Risaliti+al+2009, Zhou+Zhao+2010,
Brightman+al+2013, Jin+al+2012, Fanali+al+2013}. However, \citet{Ai+al+2011}
suggested that the relation may be flatter if $\ledd$ is higher than
$\sim1.0$. This suggestion was strengthened by \citet{Kamizasa+al+2012} using
a sample of sources with low BH masses and high $\ledd$. The relation between
$\Gamma$ and $\ledd$ in the high $\ledd$ regime is therefore still unclear, but
could hold important clues on the physical properties of near-Eddington
accretors.

In addition to the power-law continuum, other spectral features, such as
absorption, reflection and soft X-ray excess are also present in the X-ray
spectra of AGN. A significant fraction of X-ray radiation will be absorbed or
scattered by the material along the line of sight between the X-ray source and
the observer \citep[e.g.][]{Turner+al+1997, Risaliti+al+1999,
Turner+Miller+2009}. The primary effect of absorption on the X-ray spectrum is
the cut-off of the emission at low energies if absorbed by neutral material, or
resonant absorption lines and edges from ionized elements if the material is
ionized/warm gas, making it possible to measure the amount of material along
line of sight through X-ray data.  The reflection component is produced through
Compton scattering of X-ray photons into the line of sight by optically thick
material.  It will be most significant in the 5-50\,keV range with a peak at
around 30\,keV \citep[][the so-called Compton hump]{Krolik+1999}. The 6.4\,keV
fluorescent Fe K$\alpha$ line, which is expected if reflection is important
\citep{George+al+1990}, is the most prominent emission line in the X-ray
spectra of AGN \citep{Pounds+al+1990, Nandra+Pounds+1994}.  Many previous
studies found that the equivalent width (EW) of the narrow Fe K$\alpha$ line,
which is believed to be produced in the inner wall of the torus, i.e. an
optical thick dusty cloud surrounding the nucleus on parsec-scale, is
anti-correlated with the $2-10~\mathrm{keV}$ X-ray luminosity \citep[the X-ray
Baldwin effect or Iwasawa-Taniguchi effect, e.g.][]{Iwasawa+Taniguchi+1993,
Page+al+2004, Bianchi+al+2007, Shu+al+2010, Shu+al+2011, Falocco+al+2013}. The origin of this
correlation is still unclear, with one possible explanation being the decrease
of the covering factor of the torus with the X-ray luminosity
\citep{Page+al+2004, Zhou+Wang+2005, Ricci+al+2013a}. A broad Fe K$\alpha$ line
is also found in roughly $\sim45$\,per cent of AGN
\citep{Nandra+al+2007, deLaCalle+al+2010, Patrick+al+2012}, whose
archetypes include MCG-6-30-15 \citep{Tanaka+al+1995}, 1H\,0707-495
\citep{Fabian+al+2009} and others \citep[e.g.][]{Nandra+al+2007,
Patrick+al+2012}. Unlike the narrow Fe K$\alpha$ line, the broad Fe K$\alpha$
line is generally thought to originate from the inner region of the accretion
disc \citep{Fabian+al+1989}, and to be broadened by the motion induced by the
strong gravitational field of the central SMBH (alternative explanations
have also been proposed, however; e.g. \citealt{Miller+al+2008} showed that the
emergence of a broad profile can be due to complex absorption. The broad line
profile can also be well fitted with a Compton-thick wind model, see
\citealt{Tatum+al+2012}). The line profile of the broad Fe K$\alpha$ line can
be used to constrain the spin of the black hole \citep[BH;
e.g.][]{Brenneman+Reynolds+2006, Dauser+al+2010}.  All these observed X-ray
spectral features, which are likely to originate from the central region of
AGN, offer the best chance to study the immediate environment of SMBH and the
poorly understood accretion process.

AGN are traditionally divided into different types on the basis of the observed
properties, e.g. type-1 and type-2 based on the optical emission line,
radio-loud and radio-quiet depending on the radio radiation. The unification
model of AGN explains the different types by the orientation of the observer
with respect to the torus \citep{Antonucci+1993, Urry+Padovani+1995}. In this
scheme all the different types of AGN have the same central engine (SMBH) and
small scale broad-line region (BLR). The type-1 AGN is seen from inclination
angles which are not obscured by the torus, showing signature of broad-emission
lines in the optical spectra.  Conversely, the broad lines are hidden in type-2
AGN, for which the obscuration by the torus is in the line of sight, and only
the high-excitation lines produced in the narrow-line region on much larger
scales can be observed.  The detection of polarized broad-emission lines in the
optical spectra of type-2 AGN, showing the existence of BLR, is the key piece of
evidence for the unification scenario \citep{Antonucci+Miller+1985}. It is also
supported by the observed X-ray spectra, which show substantial obscuration in
type-2 galaxies.  However, results based on larger and deeper X-ray surveys
show evidences that the fraction of absorbed AGN appeared to decrease with
X-ray luminosity \citep{Ueda+al+2003, Akylas+al+2006, Hasinger+2008,
Bongiorno+al+2010, Brusa+al+2010, Merloni+al+2014, Buchner+al+2015} which seems
to be at odds with the simple unification model. Using an AGN sample selected
from the XMM-COSMOS survey, \citet{Merloni+al+2014} found that a large fraction
($\sim30$\,per cent) of the sources show discrepancies between the optical and
X-ray classifications. These results suggest that the simple
unification-by-orientation scheme is not enough to explain all the observed
phenomena, and other physical parameters may play an important role.

In this work, we describe a catalogue including 8445 point-like X-ray sources
detected in the \xxl survey in the Northern hemisphere (hereinafter: XXL-N).
We also present a study of the X-ray properties of a large AGN sample selected
from the \xxln so as to understand the X-ray Baldwin effect, the impact of
accretion rate on the structure of accretion disc and the discrepancy between
the optical and X-ray classification of AGN.  The \xxln survey, along with the
dedicated Sloan Digital Sky Survey (SDSS) ancillary follow-up programme
\citep{Menzel+al+prep}, provide one of the largest X-ray detected AGN samples
with reliable optical redshift measurements and classifications, making it
possible to investigate the X-ray properties of AGN spanning a wide region of
parameter space. The large area covered by the survey, compared to most
well-studied X-ray deep fields, also provides a unique sample of the rare AGN
population, sampling the most luminous, highest $\ledd$ objects.

This paper is structured as follows, In Section\,\ref{sec:src_detection}, we
introduce the \xxln survey. Furthermore, we describe the X-ray data reduction
as well as source detection. The optical cross-matching, data reduction and
$\mobh$ estimation are presented in Section\,\ref{sec:cata}. The full list of
the point-like X-ray sources catalogue is also included in this section. The
X-ray spectral analysis for each individual source with reliable redshift
measurement is described in Section\,\ref{sec:xray_bxa}, while in
Section\,\ref{sec:xraystack}, we present the results from X-ray spectral
stacking for the type-1 AGN sample. We present the discussion and summary of
our results in Sections\,\ref{sec:discussion} and \ref{sec:summary},
respectively.  Throughout this paper, we use the cosmological parameters $H_0 =
70\,\mathrm{km\,s^{-1}\,Mpc^{-1}}$, $\Omega_\mathrm{M} = 0.27$ and
$\Omega_\Lambda = 0.73$. All quoted errors correspond to the 68 per cent
confidence level unless specified otherwise.

\section{Source detection and data reduction}\label{sec:src_detection}

\begin{figure*}
\begin{center}
\includegraphics[width=0.9\textwidth]{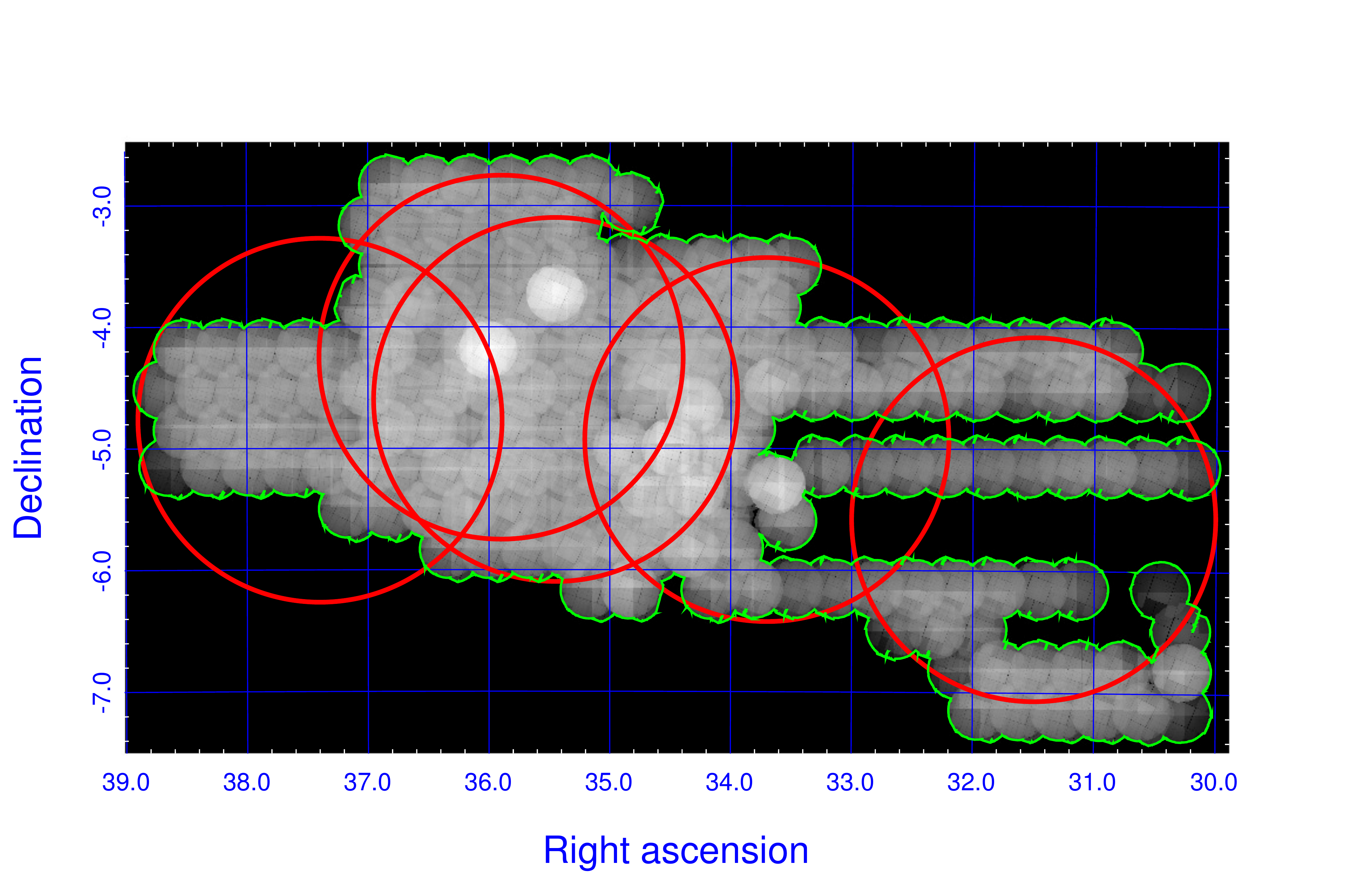}
\end{center}
\caption{Layout of the XMM observations in the equatorial XMM-XXL
field used in this paper. The distribution on sky coordinates of the
EPIC PN+MOS exposure maps of individual XMM pointings is shown. The
shading corresponds to different exposure times, with black being
zero. The green line demarcates the limits of the area covered by the
XMM data observed prior to 23 January 2012. \xxln pointings
observed after this date (total area of about $\rm 5\,deg^2$) are not
included in the analysis. This incompleteness in the X-ray data
coverage is manifested by the dark (unexposed) horizontal stripes at
e.g. $\delta \rm \approx -5.5$\,deg and $\alpha < 33.5$\,deg. The red
circles mark the positions of the five SDSS-III special spectroscopic
plates used to target X-ray sources in the field. The size of each
circle is 3\,deg in diameter, the field--of--view of the SDSS
spectroscopic plates.}
\label{fig:layout}
\end{figure*}

\begin{figure*}
\begin{center}
\includegraphics[width=1.0\columnwidth]{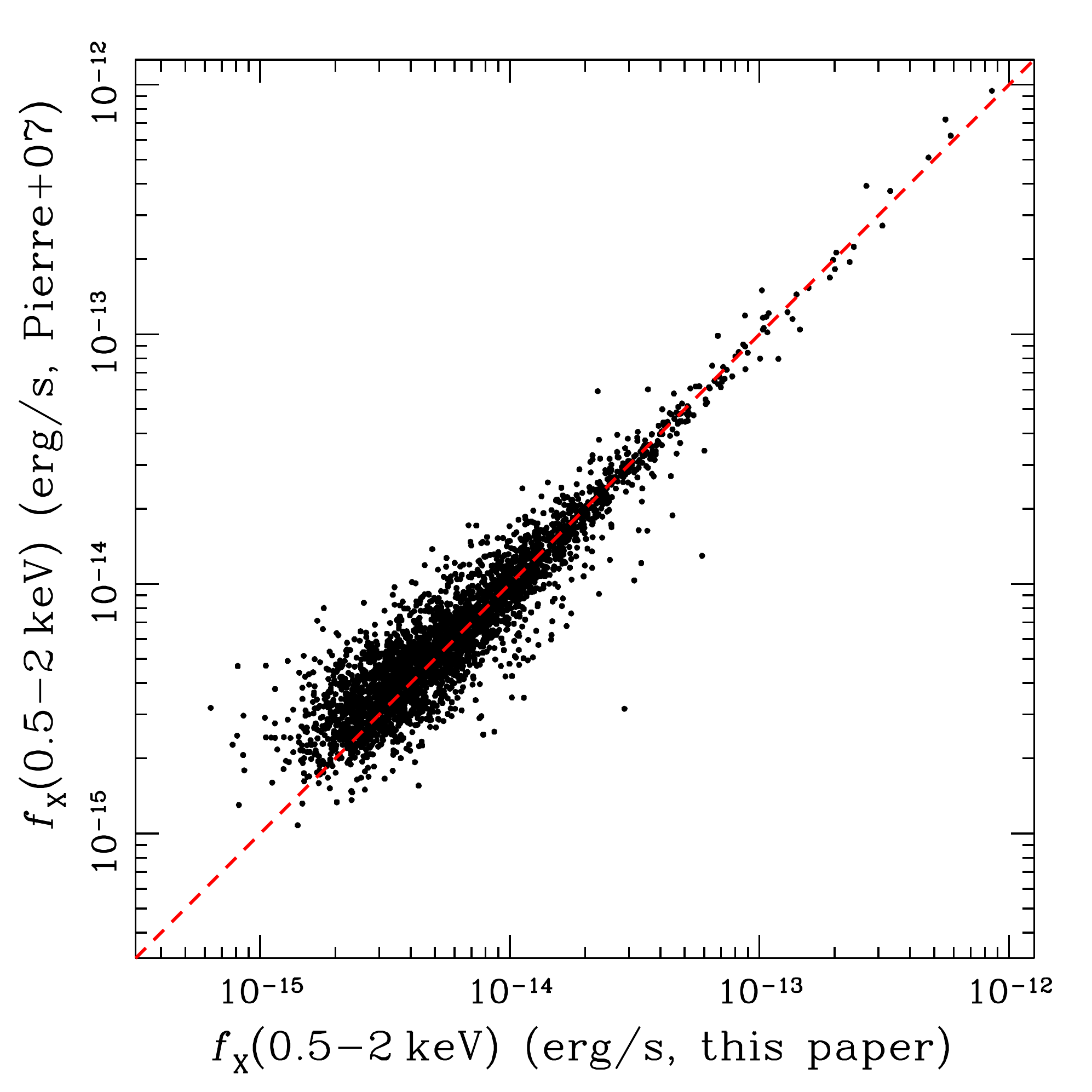}
\includegraphics[width=1.0\columnwidth]{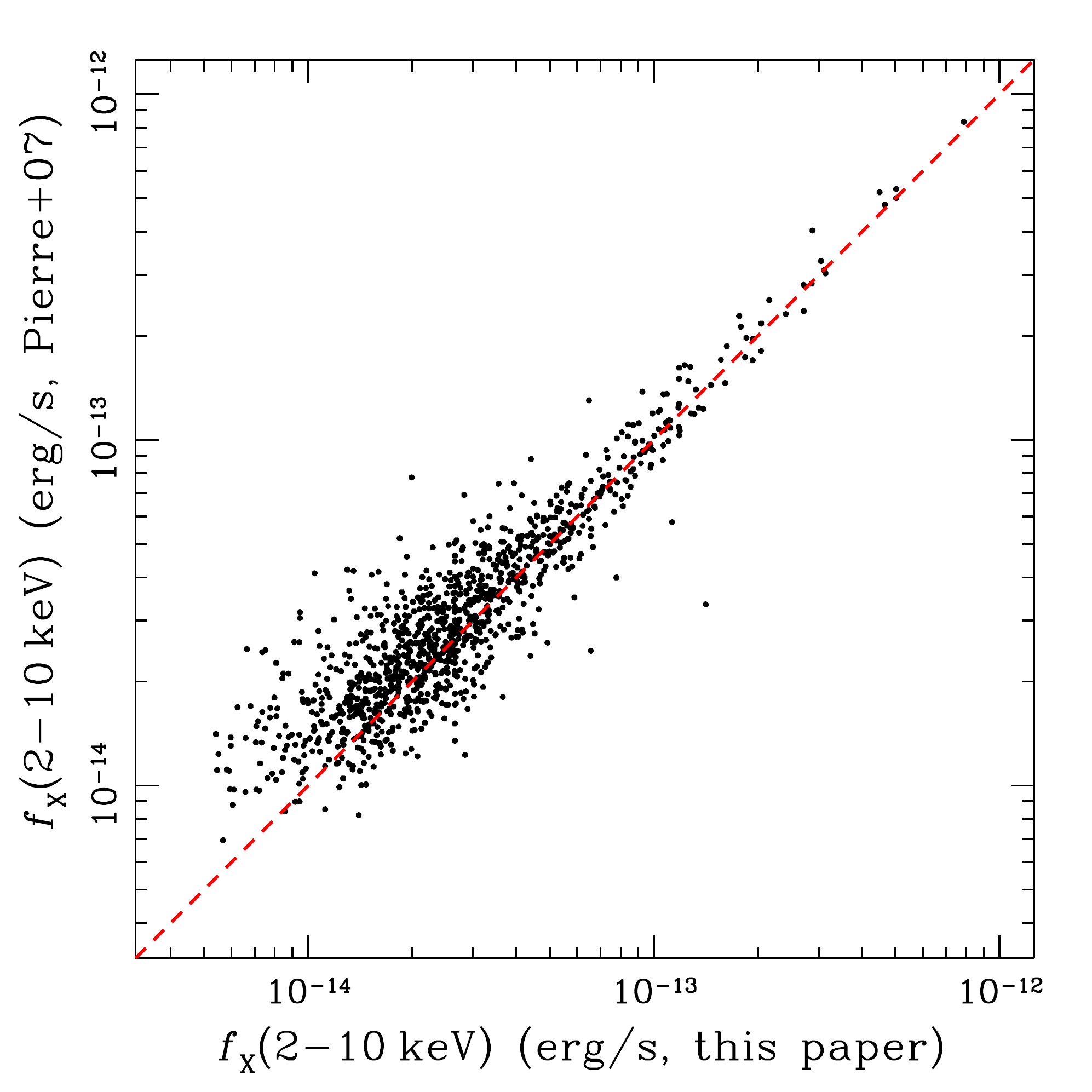}
\end{center}
\caption{Comparison of the fluxes estimated in this paper with those in the
XMM-LSS survey field \citep{Pierre+al+2015}. We used a matching radius
of 5\,arcsec to match the two X-ray source catalogues. This is results
in a total of 3369 and 1098 common sources in the soft and hard bands
respectively. Our fluxes are corrected for the different spectral
index used in this paper ($\Gamma=1.4$) and \citep[][$\Gamma=1.7$]{Pierre+al+2007}.
There is fair agreement between the two studies. The
systematic offset at the faint end is because of the Bayesian approach
adopted here to determine fluxes.}
\label{fig:compare_lss}
\end{figure*}

\begin{figure}
\begin{center}
\includegraphics[width=1.0\columnwidth]{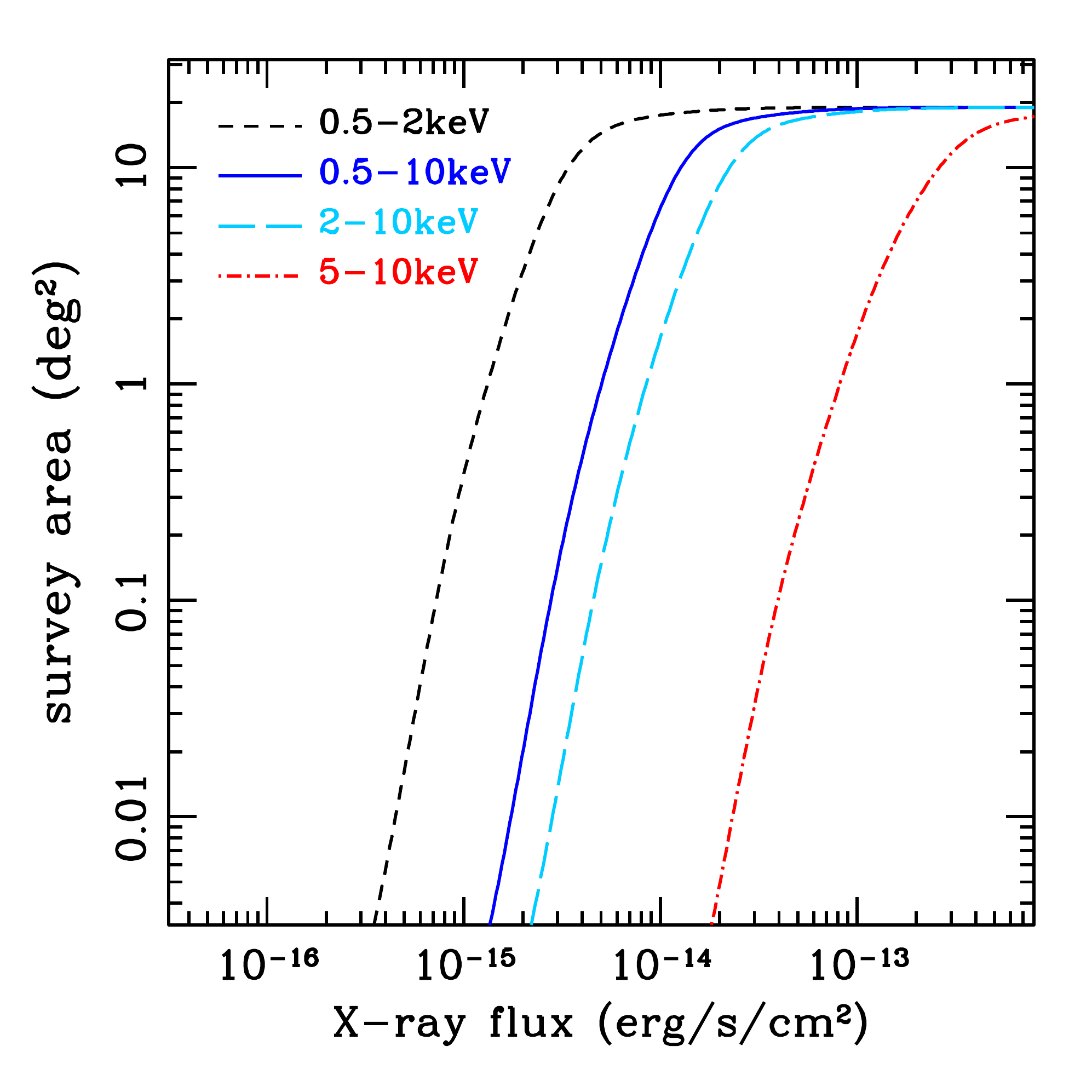}
\end{center}
\caption{Sensitivity curves in the 0.5-2, 2-10, 0.5-10 and 5-10\,keV
  spectral bands of the equatorial region of the \xxln survey
  field.}
\label{fig:sens}
\end{figure}

\begin{figure*}
\begin{center}
\includegraphics[width=1.0\columnwidth]{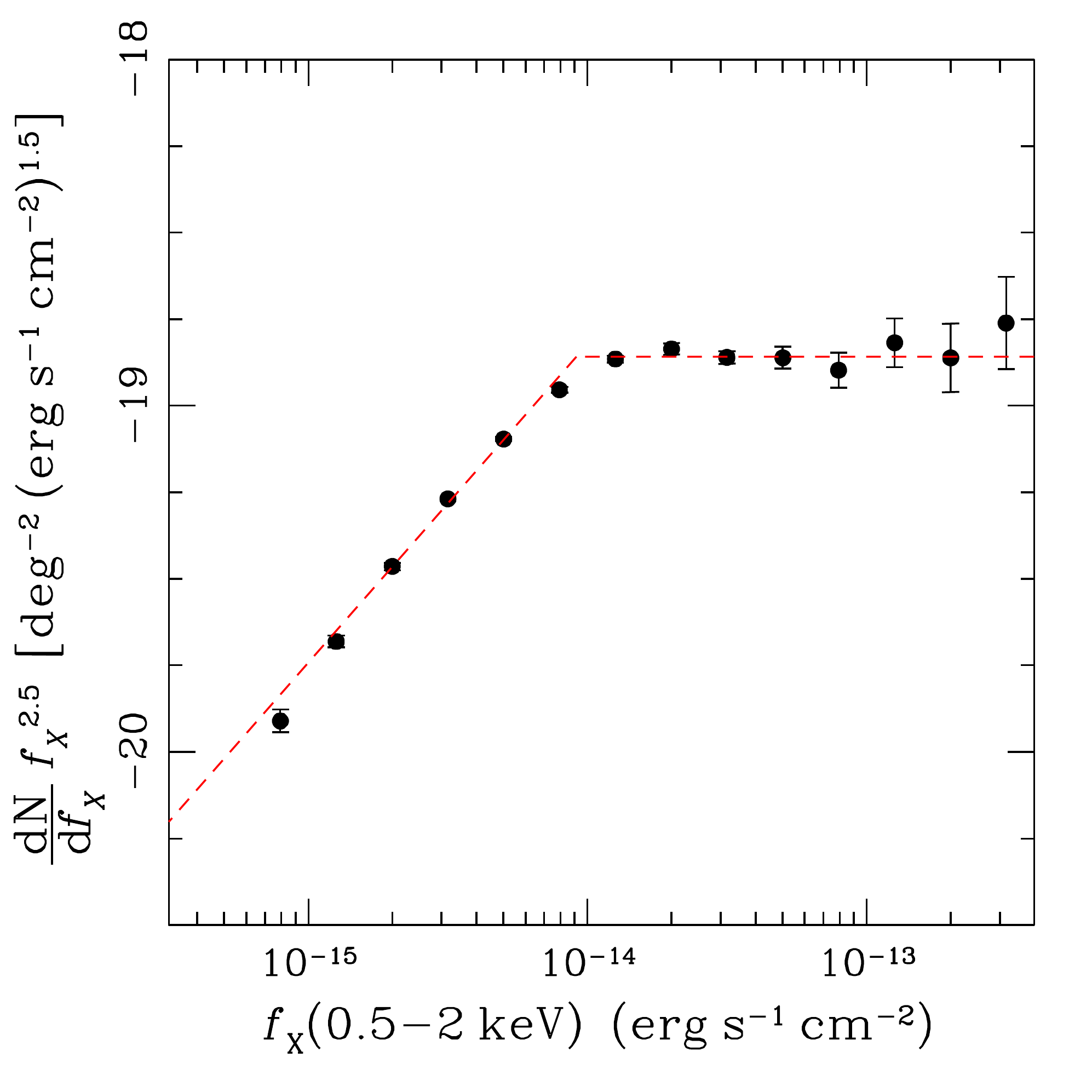}
\includegraphics[width=1.0\columnwidth]{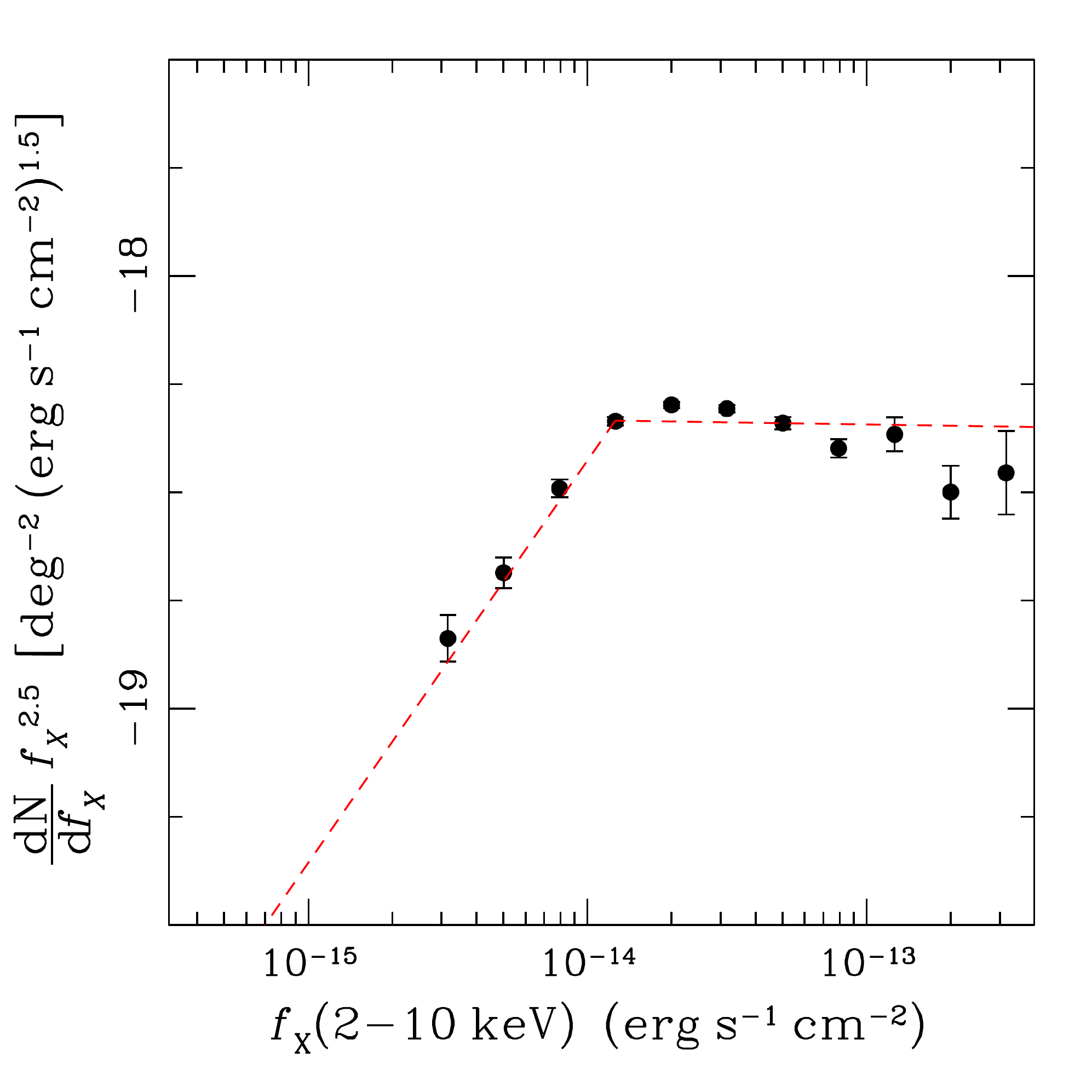}
\end{center}
\caption{\xxln differential number counts (black dots) normalized to
  the Euclidean slope in the 0.5-2 and 2-10\,keV energy range.  The (red)
  dashed line in each panel is the best-fitting relation estimated by
  \citet{Georgakakis+al+2008} using a combination of deep and shallow survey
  fields. 
}
\label{fig:dnds}
\end{figure*}

\subsection{\label{subsec:src_detection}Source detection}

The X-ray selected AGN sample used in this paper is compiled from the wide-area
and shallow \xxln survey \citep{Pierre+al+2015}. This programme covers a
total of about $\rm 50\,deg^2$ with an exposure time of about 10\,ks per XMM
pointing. The surveyed area is split into two fields nearly equal in size. In this
paper we present the source catalogue of the equatorial subregion of the \xxln,
which overlaps with the Canada-France-Hawaii Legacy Survey (CFHTLS) W1 field
and extends to about $\rm 25\,deg^2$ the area covered by the original $\rm
11\,deg^2$ XMM-LSS survey \citep{Clerc+al+2014}. The X-ray survey layout is
presented in Fig.\,\ref{fig:layout}.

The data reduction, source detection, identification of the X-ray sources with
optical counterparts and sensitivity map construction follow the methods
described in \cite{Georgakakis+Nandra+2011}. The most salient details of those
steps are outlined in this section. The XMM observations are reduced using the
{\it XMM-Newton} Science Analysis System ({\sc sas}) version 12
\citep{Gabriel+al+2004}. \xxln data observed prior to 23 January 2012 are
analysed. At that date the \xxln programme was partially complete. As a result
the final catalogue of the \xxln field presented in this paper is missing about
$\rm 5\deg^2$ worth of X-ray data. This incomplete data coverage is shown 
by the dark stripes in Fig.\,\ref{fig:layout}. Appendix \ref{tab:obs_log} lists
the individual XMM observations used in this paper. Our reduction is not
limited to the \xxln/XMM-LSS 10\,ks data only but includes deeper XMM
observations within the \xxln footprint, such as the Subaru/XMM-Newton Deep
Survey Field \citep[SXDS,][] {Ueda+al+2008}. Additionally any overlapping XMM
observations are merged using the SAS task {\sc emosaic} as explained in
\citet{Georgakakis+Nandra+2011} to increase the point source sensitivity of the
final catalogue. The first step of the data reduction process is the production
of event files from the Observation Data Files (ODF) using the {\sc epchain}
and {\sc emchain} tasks of {\sc sas} for the EPIC \citep[European Photon
Imaging Camera,][]{Struder+al+2001, Turner+al+2001} PN and MOS detectors
respectively.  Pixels along the edges of the CCDs of the PN and MOS detectors
are removed because their inclusion often results to spurious detections.
Flaring background periods are identified using a methodology similar to that
described in \cite{Nandra+al+2007}. First, the {\sc ewavlet} task of SAS
with a threshold of 5 times the local background rms is used to detect sources
in the $0.5-8$\,keV spectral band of the combined image of all EPIC detectors
available to a particular XMM observation. These sources are then masked out
from the event file before generating the background light curve. The
approximate quiescent background level is determined by calculating the count
rate at which the excess variance of the background light curve is minimum.
Periods with background count rate 2 times higher than the level where the
excess variance is minimum are excluded from the filtered event file
\citep[see][for more details]{Georgakakis+Nandra+2011}. Images and exposure
maps in celestial coordinates with pixel size of 4.35\,arcsec are constructed
in 5 energy bands, $0.5-8$ (full), $0.5-2$ (soft), $2-8$ (hard), $5-8$
(very-hard) and $7.5-12$\,keV (ultra-hard). All overlapping EPIC images are
merged prior to source detection to increase the sensitivity to point sources.
The detection algorithm is applied independently to each of the 5 spectral
bands defined above. 

The source detection is a two pass process. Source candidates are identified
using the wavelet based {\sc ewavelet} source detection task of {\sc sas} at a
low threshold of $4\sigma$ above the background, where $\sigma$ is the rms of
the background counts. For each candidate source the Poisson probability of a
random background fluctuation is estimated. This step involved the extraction
of the total counts at the position of the source and the determination of the
local background value. At each source position the counts are extracted within
an elliptical aperture that corresponds to 70 per cent of the point spread
function (PSF) encircled energy fraction (EEF), using the XMM EPIC PSF
parametrisation of \citet{Georgakakis+Nandra+2011}. The total counts at a
candidate source position, $T$, is the sum of the extracted counts from
individual EPIC cameras. For each source the local background is estimated by
first masking out all detections within 4\,arcmin of the source position using
an elliptical aperture that corresponds to the 80 per cent EEF ellipse. The
counts from individual EPIC cameras are then extracted using elliptical annuli
centred on the source with inner and outer semi-major axes of 5 and 15 pixels
(0.36 and 1.09\,arcmin) respectively. The position angle and ellipticity of the
ellipses are set equal to those of the 70 per cent EEF. The mean local
background, $B$, is then estimated by summing up the background counts from
individual EPIC cameras after scaling them down to the area of the source count
extraction region. The Poisson probability $P(T,B)$ that the extracted counts
at the source position, $T$, are a random fluctuation of the background is
calculated. We consider as sources the detections with $P(T,B)<4\times10^{-6}$.
The above methodology is optimised for the detection of point sources. Extended
X-ray sources are also present in the catalogue but our method is not designed
to identify them. For that we use the {\sc emldetect} task of {\sc sas} to
perform PSF fits to the source count distribution and to estimate the source
extent and the extent likelihood. {\sc emldetect} is applied separately to each
of the five energy bands, soft, full, hard, very-hard and ultra-hard. The free
parameters are the source count rate and source extent. The source positions
are kept fixed during the minimization process. We consider as extended those
sources with extension likelihood $>3$ in either the full or the soft spectral
bands. The {\sc emldetect} task is also used to identify potential spurious
sources in the catalogue. Point sources for which {\sc emldetect} failed to
determine a reliable fit and hence, did not estimate a detection likelihood are
marked as potentially spurious.  In the final catalogue these sources are
listed with a negative {\sc emldetect} detection likelihood. 

The source catalogues in different energy bands are merged to produce a unique
list of sources. A source detected in a particular energy band is cross-matched
with sources in other energy bands by finding the closest counterpart. The
search radius is set to the 70 per cent EEF circular radius. It should be
stressed that our detection method is geared towards point sources and
therefore the extended source catalogue should be treated with caution.
Providing a well-defined sample of X-ray selected clusters is not among our
goals.

The X-ray source positions suffer from systematic offsets because of pointings
errors of the XMM-Newton telescope at the arcsec level
\citep[e.g.][]{Watson+al+2009}. It is possible to correct for such systematic
errors by cross-matching the X-ray source positions with external catalogues
with higher astrometric accuracy. We use the {\sc eposcorr} task of {\sc sas}
to determine any systematic offsets in the astrometry of individual XMM
observations by cross-correlating the X-ray source positions with optical
sources in the SDSS-DR8 catalogue \citep{Aihara+al+2011} with magnitudes
$r<22$\,mag. The final astrometric accuracy of the X-ray source catalogue is
1.5\,arcsec ($1\sigma$ rms). 

The flux of each source in different spectral bands is estimated by assuming a
power-law X-ray spectrum with $\Gamma=1.4$, i.e. similar to the XRB. The fluxes
are corrected for Galactic absorption using the average Galactic hydrogen
column density in the direction of each XMM observation based on the HI maps of
\cite{Kalberla+al+2005}. The energy to flux conversion factors are such that
the counts from the 0.5-2, 0.5-8, 2-8, 5-8 and 7.5-12\,keV bands are
transformed to fluxes in the 0.5-2, 0.5-10, 2-10, 5-10 and 7.5-12\,keV bands
respectively.  The Bayesian methodology described by \citet{Laird+al+2009} and
\citet{Georgakakis+Nandra+2011} is adopted for the estimation of source fluxes
and the corresponding $1\sigma$ confidence intervals.  The limiting
$0.2-10$\,keV X-ray flux in the \xxln is
$\sim1.0\times10^{-15}$\,erg\,s$^{-1}$\,cm$^{-2}$.  Fig.\,\ref{fig:compare_lss}
compares our fluxes in the 0.5-2 and 2-10\,keV band with those in the XMM-LSS
X-ray source catalogue \citep{Pierre+al+2007}. A radius of 5 arcsec is used to
match the XMM-LSS catalogue, which lead to a total of 3369 and 1098 common
sources in the soft and hard bands, respectively. Overall there is good
agreement, although at the faint-end our fluxes are systematically lower than
the XMM-LSS ones. This is a direct consequence of the Bayesian approach of
estimating fluxes, which accounts for Eddington bias as well as background and
source count Poisson uncertainties. Taking these effects into account result in
flux probability distribution functions that are skewed to low fluxes
\citep[see,][]{Laird+al+2009, Georgakakis+Nandra+2011}.

The construction of the sensitivity maps follows the methodology of
\citet{Georgakakis+al+2008}. Fig.\,\ref{fig:sens} plots the sensitivity maps in
the 0.5-2, 2-10\,keV, 0.5-10, and 5-10\,keV energy bands. The construction of
the X-ray number counts provides an independent consistency check of the
overall data reduction process, including source detection, flux estimation and
sensitivity map generation. Fig.\,\ref{fig:dnds} presents the differential
X-ray number counts in the 0.5-2, 2-10\,keV. For comparison that figure also
plots the best-fitting differential count relations of \citet{Georgakakis+al+2008}
based on a combination of Chandra deep and shallow survey fields. There is
encouraging agreement indicating that the X-ray source detection and
sensitivity map generation are robust and consistent. 

\subsection{\label{subsec:data_reduction}Extraction of X-ray spectra}

The source spectra were extracted from circular source-centred regions, of
which the radii were chosen to maximize the signal-to-noise over the 2-10 keV
band via the SAS task {\sc eregionanalyse}. To select the background regions,
we first masked out all the source regions in each image. In the case of
the EPIC MOS camera, concentric annulus with inner radii of 40\,arcsec and
outer radii of 110\,arcsec is selected if it is placed on the same CCD chip as
the corresponding source. Otherwise, circular region of 90 arcsec radii, which
is placed on the same CCD chip with roughly the same off-axis angle as the
corresponding source, is selected. In the case of EPIC PN camera, the
background events were extracted from source-free regions using circle of 90
arcsec radius at the same CCD read-out column as the corresponding source
position. If this was not possible, e.g. the source might lie close to the CCD
gaps, then a circle of 90 arcsec located at the same CCD chip was selected.
X-ray events with pattern$~\leq12$ for MOS camera and $\leq4$ for PN camera
were used to extract the X-ray spectra. The redistribution matrix and ancillary
file were generated using the SAS task {\sc rmfgen} and {\sc arfgen},
respectively. For each source, to increase the S/N, the two MOS spectra in each
observation were combined to form a single MOS spectrum if both the MOS1 and
MOS2 spectra were available.

\section{Optical counterpart identification and the source catalogue}\label{sec:cata}

%%%%%%%%%%%%%%%%%%%%%%%%%%
%%% The catalogue
%%%%%%%%%%%%%%%%%%%%%%%%%%

In total, there are 8445 point-like X-ray sources detected in the \xxln field.
The source detection threshold is 4$\sigma$ above the background and the
Poisson probability is less than $4\times10^{-6}$ (see
Section\,\ref{subsec:src_detection}).  In this section, we briefly describe the
optical data analysis of the sources which have Sloan Digital Sky Survey (SDSS)
observations \citep[more details can be found in][]{Menzel+al+prep}.

\subsection{\label{subsec:opt_obs}Cross-matching with optical data}

X-ray sources in the \xxln survey were matched to the SDSS-DR8 photometric
catalogue \citep{Aihara+al+2011} following the Likelihood Ratio method
\citep{Sutherland+Saunders+1992} as implemented in
\citet{Georgakakis+Nandra+2011}. For the positional accuracy of the X-ray
sources, we adopt a Gaussian distribution with a standard deviation of
1.5\,arcsec (see above). A radius of 4\,arcsec around the X-ray position is
used to search for optical counterparts. We chose to use a likelihood threshold
for secure counterparts $LR_{\rm limit} = 1.5$. At this cut-off, 48 per cent
(4075/8445) of the X-ray point sources have SDSS counterparts with an
estimated spurious identification rate of about 6 per cent.

The full source catalogue is made available to the public
\footnote{\label{url_cata}\url{http://www.mpe.mpg.de/XraySurveys/XMM-XXL/XMM-XXL/ADD-ONs/Xray_data/xxl_xray_detected_source_2016MAR30.fits}}
in FITS (Flexible Image Transport System) format. The information
included in that file is listed in Appendix\,\ref{tab:tab_col}.

\subsection{\label{subsec:opt_info}Redshift measurements and optical
classifications}

\begin{table*}
\bgroup
\setlength\tabcolsep{8pt}
\begin{center}
\caption{\label{tab:fraction_types}The fraction of different types of sources
in our spectroscopic AGN sample. Sources are classified as BLAGN1 (or optical type-1)
if one of the optical lines, i.e. H$\beta$, Mg\,\textsc{ii}, C\,\textsc{iii]}
or C\,\textsc{iv}, have FWHM larger than $1000~{\rm km~s^{-1}}$. NLAGN2 (or optical
type-2) are objects with FWHMs of all optical emission lines less than $1000~{\rm km~s^{-1}}$.
Objects with no significantly detected optical emission lines are defined as NLAGN2cand.
The eAGN-ALG are sources whose optical
spectra show absorption lines and a few narrow emission lines which are not significant,
while eAGN-SFG are objects with optical emission lines from young stars. The remaining
sources (defined as unknown) cannot be reliably classified due to the low S/N of their
optical spectra.}

\begin{tabular}{|l|c|c|c|c|c|c|c|c|c|}
\hline\hline
\raisebox{-2.5mm}{Classification} & & Optical type-1 & & \multicolumn{4}{|c|}{Optical type-2} & &\raisebox{-2.5mm}{Unknown} \\[-0.5mm]\cline{3-3}\cline{5-8}\\[-2.3mm]
			  & &	BLAGN1	    & & NLAGN2	& NLAGN2cand & eAGN-SFG & eAGN-ALG & & \\\hline
Fractions	  & &		0.70    & &  0.10     &     0.11   &   0.03   &   0.04 & & 0.02 \\\hline
\end{tabular}
\end{center}
\egroup
\end{table*}

Among the 8445 X-ray detected sources, a total of 3042 sources, in the optical
magnitude range $15.0<r_{AB}<22.5$, were followed up by the BOSS spectrograph
\citep{Menzel+al+prep}. The redshift measurements as well as the optical
classifications are extensively discussed in \citet{Menzel+al+prep}. Here we
only outline the main procedures. The optical spectra are first analysed by the
BOSS {\tt spec1d} pipeline, which assigns a redshift and a classification using
the $\chi^2$-minimization method \citep{Stoughton+al+2002, Bolton+al+2012}.
However, the BOSS pipeline may fail to measure the redshift correctly for some
sources, in particular for those of low signal to noise (S/N) or very high
redshift (see \citealt{Menzel+al+prep} for a list of reasons). Thus a dedicated
{\it visual} inspection is applied to $\sim1200$ low S/N spectra in our data.
At the end of the process, there are in total 2570 extragalactic sources which
have reliable redshift measurements. Those sources are then classified based on
their optical emission lines. Generally, the sources are classified as
broad-line AGN (optical type-1) and narrow-line AGN ( optical type-2) based on
the full width at half maximum (FWHM) of their optical emission lines, i.e.
H$\beta$, Mg\,\textsc{ii}, C\,\textsc{iii]} or C\,\textsc{iv}. In addition, sources that do not show any
AGN-driven emission lines in their optical spectra are referred to as elusive
AGN (eAGN). Sources are classified as BLAGN1 (or optical type-1) if one of the
optical lines have FWHM larger than $1000~{\rm km~s^{-1}}$. NLAGN2 (or optical
type-2) are objects with FWHMs of all the H$\beta$, Mg\,\textsc{ii}, C\,\textsc{iii]} or C\,\textsc{iv} emission
lines less than $1000~{\rm km~s^{-1}}$. However, in some cases, due to strong
host galaxy continuum contribution or very low S/N ratio spectra, the optical
emission lines of some sources are not significantly detected. These objects
are defined as NLAGN2cand. The elusive AGN are further classified as eAGN-ALG
and eAGN-SFG depending on their optical spectral properties. The eAGN-ALG are
sources whose optical spectra show absorption lines and a few narrow emission
lines which are not significant, while eAGN-SFG are objects with optical
emission lines from young stars. In this work, the optical type-2 AGN include
all the NLAGN2, NLAGN2cand and the elusive AGN. In total, we have 1787 type-1
AGN (70 per cent), 726 type-2 AGN (28 per cent, NLAGN2: 10 per cent,
NLAGN2cand: 11 per cent, eAGN-SFG: 3 per cent, eAGN-ALG: 4 per cent). The
remaining 57 sources (2 per cent) cannot be reliably classified due to the low
S/N of their optical spectra. The fraction of different types of sources
in our sample can be found in Table\,\ref{tab:fraction_types}.

\subsection{\label{subsec:mbh_edd}$\mobh$ estimation}

To estimate single-epoch virial BH masses from continuum luminosities and
broad-line widths \citep[e.g][]{Shen+2013}, we perform spectral fits to the
BOSS spectroscopy of the type-1 objects in our sample. The spectral fitting
procedure is similar to the global fitting scheme described in
\citet{Shen+Liu+2012}.  We fit a global pseudo-continuum model (including a
power-law component, broad Fe\,\textsc{ii} emission, and a low-order polynomial to
account for spectral curvature due to reddening) to several continuum windows
free of major emission lines (other than the copious Fe\,\textsc{ii} emission). For the
UV Fe\,\textsc{ii} template, we use the \citet{Vestergaard+Wilkes+2001} template
($1000-2200$\,\AA), and the modified template by \citet{Salviander+al+2007} for
the Mg\,\textsc{ii} region ($2200-3090$\,\AA); we augment the $3090-3500$\,\AA\ region
using the template derived by \citet{Tsuzuki+al+2006}. For the optical Fe\,\textsc{ii}
template ($3686-7484 $\,\AA) we use the one provided by
\citet{Boroson+Green+1992}.

The best-fitting pseudo-continuum model is subtracted from the spectrum, leaving
the emission-line-only spectrum. We then fit the several major broad lines
(e.g., H$\beta$, Mg\,\textsc{ii}, C\,\textsc{iv}) that are commonly used for single-epoch virial BH
mass estimators with multiple Gaussians (in logarithmic wavelength space). The
details of the fitting parameters and constraints can be found in
\citet{Shen+Liu+2012}. We estimate the continuum luminosities and the
broad-line FWHMs from the model fits. Finally, we use a Monte Carlo method to
estimate the uncertainties in the measured spectral quantities
\citep[e.g.][]{Shen+al+2008, Shen+al+2011}, in which we perturb the original
spectrum with the reported flux density errors, and repeat the same fitting
procedure on 50 trials of mock spectra. The 1$\sigma$ errors on the measured
spectral quantities are estimated as the semi-amplitude of the range defined by
the 16 per cent and 84 per cent percentiles in the distribution from the 50
trials.

With the measured continuum luminosities and broad-line FWHMs, we use several
single-epoch virial mass estimators to estimate BH masses. The fiducial mass
recipes we adopt are: H$\beta$ at $z<0.9$ \citet{Vestergaard+Peterson+2006};
Mg\,\textsc{ii} at $0.9<z<2.2$ \citep{Shen+al+2011}; C\,\textsc{iv} at $z>2.2$
\citep{Vestergaard+Peterson+2006}. These redshift divisions differ slightly
from \citet{Shen+al+2011} for DR7 quasars, given the slight difference in the
spectral coverage between SDSS-I/II and SDSS-III.

\begin{figure} 
    \begin{center}
    \includegraphics[width=\columnwidth]{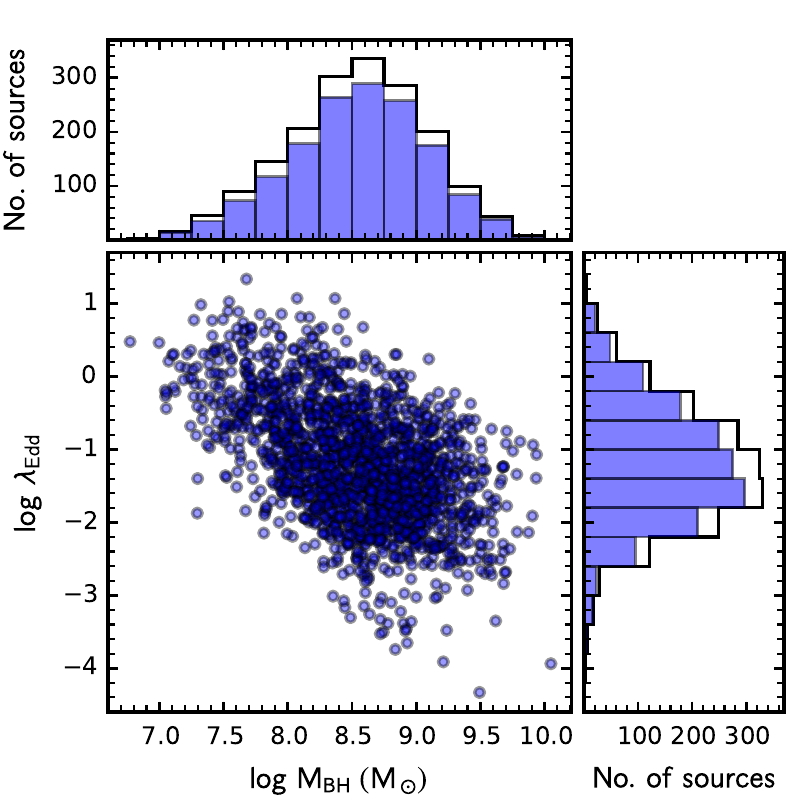}
    \caption{\label{fig:mbh_edd}The central panel shows the distribution in the
    $\log\mobh-\log\ledd$ of the 1787 type-1 sources in the spectroscopic
    sample. The black empty histograms in the top and right panels show the
    distributions of the $\log\mobh$ and $\log\ledd$ for all the type-1 AGN,
    respectively. The blue filled histograms are for the type-1 AGN sample used
    in the X-ray spectral stacking (see Section \ref{sec:xraystack}).}
    \end{center} 
\end{figure}

The measurement errors on these virial BH masses are propagated from
measurement errors in continuum luminosity and broad-line FWHM, which could be
substantial for noisy spectrum. In addition, there is a substantial systematic
uncertainty of $\sim 0.4-0.5$ dex in these single-epoch virial BH masses
\citep[see extensive discussions in][]{Shen+2013}, and we caution on the
consequences of the large systematic uncertainties in these BH mass estimates.
The distribution of the $\mobh$ is shown in the top panel (black empty
histogram, the blue filled histograms are for the type-1 AGN sample used in the
X-ray spectral stacking, see Section \ref{sec:xraystack}) of
Fig.\,\ref{fig:mbh_edd}.

\begin{figure*}
\begin{center}
  \includegraphics[width=0.48\textwidth]{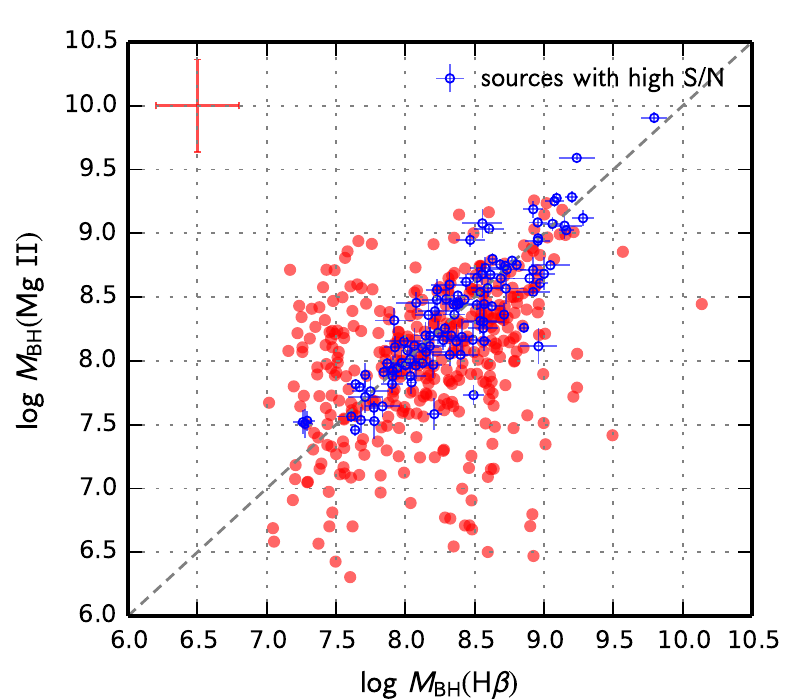}
  \includegraphics[width=0.48\textwidth]{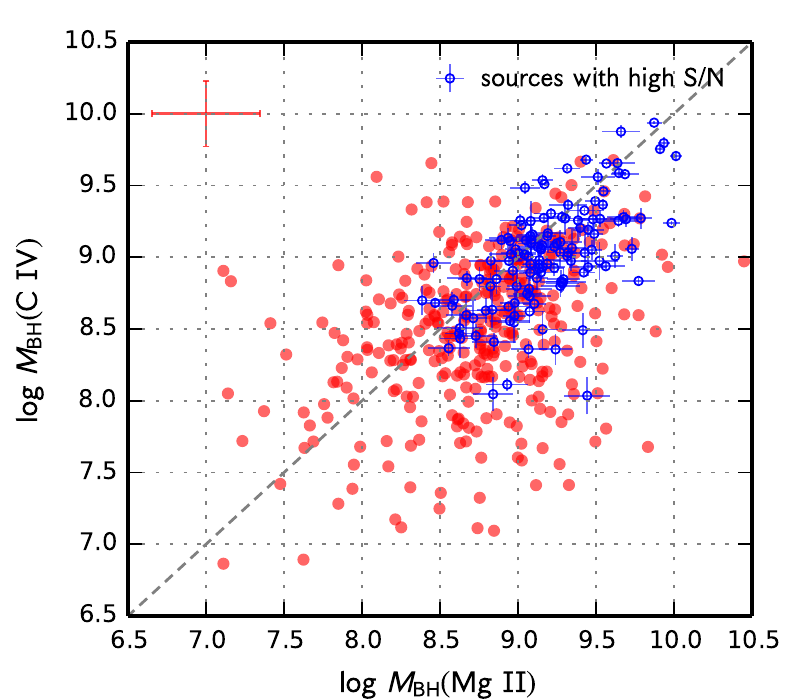}
  \caption{\label{fig:mbh_comparison}Left: comparison of $\mobh$ estimated using Mg\,\textsc{ii} to $\mobh$
  obtained using H$\beta$. Right: $\mobh$ calculated using C\,\textsc{iv} against
  $\mobh$ obtained using Mg\,\textsc{ii}. Blue circles with error bars are sources
  with higher S/N, i.e. the uncertainties of the measured $\mobh$ are less than 0.15.
  The remaining sources with lower S/N are indicated with red points. The $\mobh$ of
  these sources have uncertainties of $\sim 0.3-0.4$\,dexs, as shown by the typical
  error bars in the upper-left of each panel.}
\end{center}
\end{figure*}

Comparisons of SE BH masses using different emission lines have been
performed in several previous studies (e.g., \citealt{Shen+al+2008,
Shen+al+2011, Shen+Liu+2012}, \citealt{Shen+2013} and references therein).
These studies revealed that SE BH masses estimates using different emission
lines, when adopting the fiducial SE mass formula, are generally consistent
with each other with negligible systematic offsets and scatters that are lower
than the systematic uncertainties of these mass estimates. This ensures there
is no systematic bias in using different lines to estimate BH masses at
different redshifts in this study. As shown in Fig.\,\ref{fig:mbh_comparison}
(left: $\mobh$ calculated using Mg\,\textsc{ii} against
$\mobh$ obtained using H$\beta$, right: $\mobh$ estimated using C\,\textsc{iv} against
$\mobh$ obtained using Mg\,\textsc{ii}),
our results are generally in agreement with previous studies. The large scatter found
in Fig.\,\ref{fig:mbh_comparison} is mainly due to the low S/N of the data, especially for the
correlation between the BH mass estimated with H$\beta$ and Mg\,\textsc{ii}. This is clearly
illustrated in the left panel of Fig.\,\ref{fig:mbh_comparison}. The $\mobh$(H$\beta$) is consistent with
the BH mass obtained using Mg\,\textsc{ii} if we limit the data to sources with high S/N.
However, as shown in the right panel of Fig.\,\ref{fig:mbh_comparison}, the scatter is still
large even for sources have well determined \ion{Mg}{ii}- and \ion{C}{iv}-based BH mass estimations.
Previous studies have also shown that the C\,\textsc{iv}-based BH mass
estimates have a larger scatter around H$\beta$-based mass estimates than that
for Mg\,\textsc{ii}. We thus note that C\,\textsc{iv} may be a less reliable line in estimating BH
masses compared to H$\beta$ and Mg\,\textsc{ii} (see discussion in
\citealt{Shen+2013}). 

\section{Properties of AGN with redshift measurements}\label{sec:xray_bxa}
%%%Properties of the sources in the field

\subsection{\label{sec:xray_prop}X-ray properties}

\begin{figure} 
    \begin{center}
    \includegraphics[width=\columnwidth]{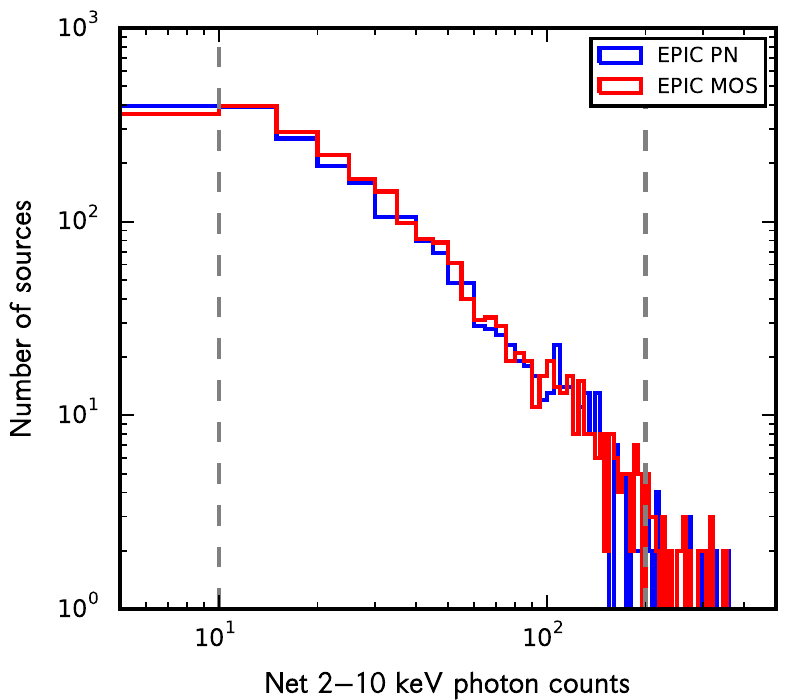}
    \caption{\label{fig:pho_dist}Distribution of the rest-frame
        2-10 keV net photon counts for the all the sources which have
        reliable redshift measurements. The vertical dashed lines
    indicate the photon count boundaries of the AGN sample presented
    in the Section\,\ref{sec:xraystack}.
    }
\end{center} 
\end{figure}

In this work, a total of 2512\footnote{Due to calibration issues, one source
(UXID: N\_121\_7) is excluded from the X-ray analysis.} sources, which have
reliable redshift measurements and optical classifications, are used to study
the X-ray properties of AGN in the {\xxln}. The background-subtracted
rest-frame $2-10$\,keV photon counts distribution of the sources in our sample
is shown in Fig.\,\ref{fig:pho_dist} (blue for PN and red for MOS). We adopted
the Bayesian X-ray Analysis software \citep[BXA,][]{Buchner+al+2014} to analyse
the X-ray spectrum of each individual source. The detail methodology of the
Bayesian approach is present in \citet{Buchner+al+2014}. The best model for
describing the X-ray spectra of a large sample of AGN found in that paper, i.e.
is used to fit the X-ray spectra of the sources
in our sample. This model consists of three main components: a) the \textsc{bntorus}
model \citep{Brightman+Nandra+2011}, which includes an intrinsic power-law
component modified by photoelectric absorption and Compton scattering from
absorption material with toroidal geometry, is used to model the power-law
continuum as well as the absorption and Compton scattering feature observed in
X-ray spectra of AGN; b) the \textsc{pexmon} model \citep{Nandra+al+2007} is included to account for the
reflection component which is also absorbed by the same material as the
intrinsic power-law continuum; c) a soft scattering component, which is
approximated with an unabsorbed power-law, is also added to model the soft
X-ray excess (soft scattering) found in type-1 (type-2) AGN. The opening angle
and the inclination of the \textsc{bntorus} model are fixed at $45^\circ$ and
$85^\circ$, respectively. The photon index of the \textsc{pexmon} model, as well as the
scattering components, are linked to the power-law component in the {\sc
bntorus} model. The inclination parameter in the \textsc{pexmon} model is fixed at
$60^\circ$. The normalization of this component is modelled relative to that
of the {\sc bntorus} model and is allowed to vary between 0.1 and 100. The same
is applied to the normalization of the soft scattering component, but with an
upper limit of 0.1 and a lower limit of $10^{-9}$.

\citet{Liu+Li+2015} have recently shown that the \textsc{bntorus} model will over predict
the soft-Xray emission in heavily obscured objects, comparing with the results
obtained from the \textsc{mytorus} model \citep{Murphy+Yaqoob+2009} and
the model built by \citet{Liu+Li+2014}. Using the data observed with
the \textit{Nuclear Spectroscopic Telescope Array} \citep[\textit{NuStar}][]{Harrison+al+2013},
there have been many studies that fit the spectra with both \textsc{mytorus}
and \textsc{bntorus} models. They found no evidence that
one model gives significantly different parameters than the other, except for sources
with $N_\mathrm{H}$ higher than $10^{25}\,\mathrm{cm^{-2}}$ or extreme opening angles found with \textsc{bntorus}
\citep[e.g.][]{Bauer+al+2015, Arevalo+al+2014}.
\citet{Brightman+al+2015} also conducted comparisons of \textsc{mytorus} and \textsc{pexrav} \citep{Magdziarz+Zdziarski+1995}  models
with \textsc{bntorus} using a sample of AGN observed by \textit{NuStar}.
They again found that the parameters given by the two models are consistent with
each other, when both models are valid. The reason for the agreement,
despite the differing reflection strength predicted by \textsc{mytorus} and \textsc{bntorus},
is that the difference is only large in the softer energies where typically
scattered/thermal emission dominates over the reflection. In this work, the majority
of the sources in the AGN sample have column density less than $10^{25}\,\mathrm{cm^{-2}}$ (see below) and
a soft scattering component is included to model the scattered emission for heavily obscured
objects, thus the \textsc{bntorus} model is still appropriate.

We jointly fit the PN and MOS spectra for each source, if both are available.
A MOS/PN normalization factor is thus included to allow the normalization of
the MOS spectra to vary.  We have in total 6 free parameters, the photon index
$\Gamma$ of the intrinsic power-law, the column density $N_{\rm H}$, the MOS/PN
normalization factor, the normalization of the torus, Compton reflection and
scattering components. To apply the BXA method, priors need to be assigned for
each of these parameters. Following \citet{Buchner+al+2014}, we applied
log-uniform priors on all normalization and column density
parameters, while Gaussian priors are used for the photon index $\Gamma$ (with
the mean of 1.95 and standard deviation of 0.15) and the MOS/PN normalization
factor (with the mean of 1.0 and standard deviation of 0.25 based on the
results of a simple power-law fitting) parameters. In addition, models for
the {\it XMM-Newton} EPIC PN and MOS background spectra constructed by
\citet{Maggi+al+2014}, which include an empirical instrumental background
component developed by \citet{Sturm+2012} and astrophysical background
component of \citet{Kuntz+Snowden+2010}, are adopted in order to apply the BXA
method.

\begin{figure} 
    \begin{center}
    \includegraphics[width=\columnwidth]{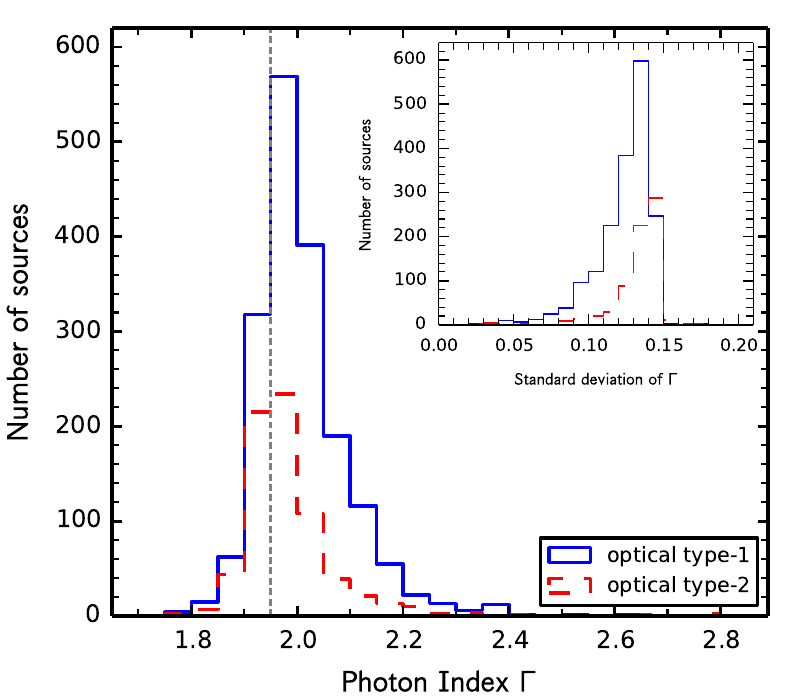}
    \caption{\label{fig:phoind_dist} Distribution of photon index for type-1
    (solid blue) and type-2 (dashed red) AGN.  For each of the source, the
    median value of the posterior distribution is used.  The mean value,
    $\Gamma=1.95$, for the photon index prior is indicated with vertical dashed
    line. Inset: Distribution of the standard deviation of $\Gamma$ for type-1
    (solid blue) and type-2 (dashed red) AGN.}
\end{center} 
\end{figure}

The spectra are fitted in the observed $0.5-8$\,keV energy range. Hereafter the
quoted value of each parameter is the median of its posterior distribution
unless specified otherwise. As shown in Fig.\,\ref{fig:phoind_dist},
the distribution of the photon index parameter has a similar shape as the input priors,
though the mean photon indices for both type-1
and type-2 AGN ($\Gamma\sim2.0$) are higher than the mean of the Gaussian prior.
The standard deviations of the photon index are also
peaked at values close to $\sim0.15$ (see the inset in Fig.\,\ref{fig:phoind_dist}).
This may be an indication that the posterior distributions of the photon index
are driven by the priors. To test this, the photon index posteriors are re-analysed
under a flat prior, using the methodology described in \citet[][see their Appendix A]{Buchner+al+2015}.
We find that, assuming a flat prior, the intrinsic distribution of the posteriors
can be well described by a Gaussian of mean 2.14 (2.10) and
standard deviation 0.22 (0.33) for type-1 (type-2) AGN, consistent with the result
obtained by assuming a Gaussian prior. Thus, the result shown in Fig.\,\ref{fig:phoind_dist}
is {\it not} driven by the prior, and is robust against the choice of
prior. The asymmetry of the
photon index distribution is due to sources which have very soft and steep
X-ray spectra, e.g. sources with strong soft X-ray excess.

\begin{figure} 
    \begin{center}
    \includegraphics[width=\columnwidth]{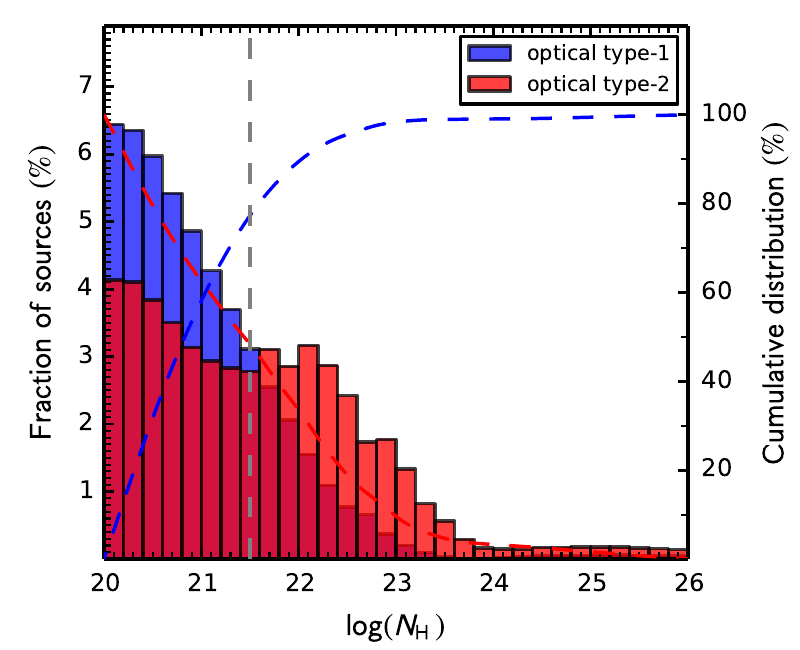}
    \caption{\label{fig:nh_dist}Normalized distribution of the column density
    parameter for sources with photon counts larger than 10 in either the PN or
    MOS spectra. For each source, the full posterior distribution is used.
    Optical classified type-1 and type-2 AGN are shown in blue and red
    histogram, respectively. The blue dashed line shows the cumulative
    distribution $C(\log{N_{\rm H}})$ for type-1 AGN and the red dashed line
    shows the negative cumulative distribution $1-C(\log{N_{\rm H}})$ for
    type-2 AGN.  The vertical dashed line indicates the $\log{N_{\rm H}}$ value
    we used to define the X-ray obscured/unobscured samples (see
    Section\,\ref{sec:x_opt_class}).}
\end{center} 
\end{figure}

For the rest of the parameters with log-uniform priors and the MOS/PN
normalization factor parameter, though with large uncertainties, the fitted
values also provide interesting information (see
Section\,\ref{sec:x_opt_class}). Fig.\,\ref{fig:nh_dist} shows the
distributions of the column density for sources with more than 10 photon counts
in either the PN or MOS spectra (blue: type-1 AGN; red: type-2 AGN). The
cumulative \nh distribution of type-1 AGN (blue curve, fraction of sources with
column density {\it smaller} than a given value) and the reversed cumulative
\nh distribution of type-2 AGN (red curve, the fraction of sources with column
density {\it larger} than a given value) are also shown in
Fig.\,\ref{fig:nh_dist}.  The full posterior distribution of the column density
parameter is used to generate this plot.  Generally, the optically selected
type-2 AGN have a different distribution from the type-1 AGN, i.e. with more
sources having column density higher than $10^{22}$\,cm$^{-2}$. It also clearly
shows that $\sim40$\,per cent of type-1 (type-2) AGN have X-ray measured line
of sight column density larger (lower) than $10^{21}$\,cm$^{-2}$. Our
results, with a larger spectroscopic AGN sample, it is consistent with the
conclusion found in \citet[][see also
\citealt{Trouille+al+2009}]{Merloni+al+2014} using an AGN sample selected from
the \xmmcosmos survey, though with a slightly smaller intersection value (i.e.
$\sim10^{20.9}~{\rm cm^{-2}}$ in our sample and $\sim10^{21.1}~{\rm cm^{-2}}$
in their spectroscopic sample) of the two cumulative distribution.

\begin{figure} 
    \begin{center}
    \includegraphics[width=\columnwidth]{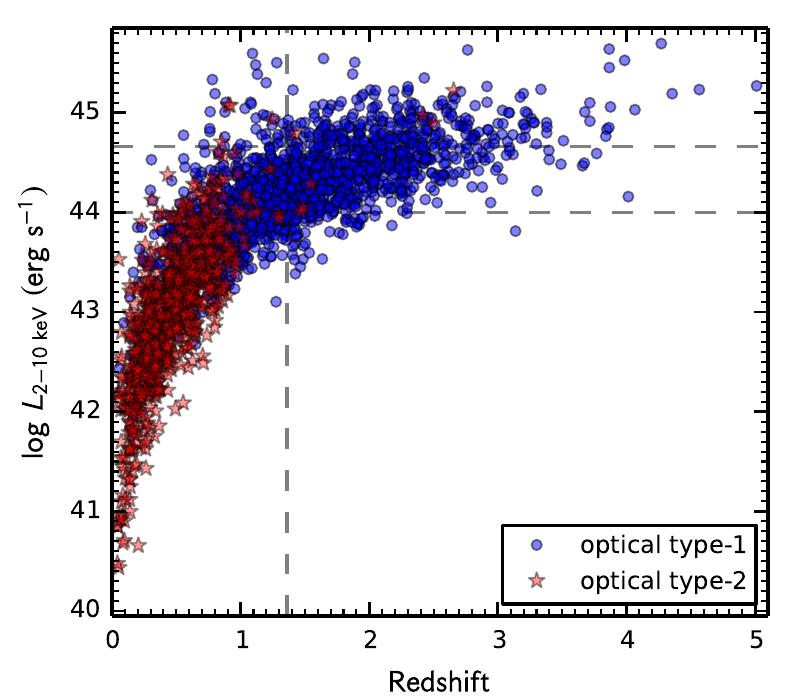}
    \caption{\label{fig:lum_red_dist}Distribution of sources in the rest-frame
    intrinsic 2-10 keV luminosity-redshift plane for type-1 (blue points) and
    type-2 (red stars) AGN. The luminosity and redshift boundaries for the
    type-1 redshift-luminosity sub-samples (see Section\,\ref{sec:sub_sample})
    are indicated by the vertical and horizontal dashed lines.}
    \end{center} 
\end{figure}

To derive the rest-frame $2-10$\,keV intrinsic X-ray luminosity
distribution, both the Galactic and host galaxy column densities are fixed at
zero.  The posterior distribution for the X-ray luminosity of each source is
then calculated using the \textsc {bntorus+pexmon+scattering} model.
Fig.\,\ref{fig:lum_red_dist} shows the distribution of the sources in the
$\log\lxray-z$ plane for type-1 (blue points) and type-2 (red stars) AGN.

The median values of the posterior distributions for the parameters are
included in Table\,\ref{tab:xray_cata}. In Table\,\ref{tab:xray_cata}, we also
present the redshift, optical classification, $\lxray$ as well as the
rest-frame 2-10~keV net photon counts for each source. The full table can be
found online, in which the 3rd, 16th, 84th and 97th percentiles of each
parameter are also included.

\begin{table*}
%\begin{minipage}{120mm}
\begin{center}
\caption{\label{tab:xray_cata}Properties of sources with reliable redshift measurement and optical classification}
\begin{tabular}{|l|r|l|c|c|c|c|c|c|c|c|}
\hline\hline
  \multicolumn{1}{|c|}{UXID} &
  \multicolumn{1}{c|}{$z$} &
  \multicolumn{1}{c|}{classification} &
  \multicolumn{1}{c|}{$N_{\rm H}$} &
  \multicolumn{1}{c|}{$\Gamma$} &
  \multicolumn{1}{c|}{normalization} &
  \multicolumn{1}{c|}{reflection component} &
  \multicolumn{1}{c|}{soft scattering} &
  \multicolumn{1}{c|}{$\lxray$}&
  \multicolumn{1}{c|}{cts PN} &
  \multicolumn{1}{c|}{cts MOS}\\
  \multicolumn{1}{|c|}{(1)} &
  \multicolumn{1}{c|}{(2)} &
  \multicolumn{1}{c|}{(3)} &
  \multicolumn{1}{c|}{(4)} &
  \multicolumn{1}{c|}{(5)} &
  \multicolumn{1}{c|}{(6)} &
  \multicolumn{1}{c|}{(7)} &
  \multicolumn{1}{c|}{(8)} &
  \multicolumn{1}{c|}{(9)}&
  \multicolumn{1}{c|}{(10)} &
  \multicolumn{1}{c|}{(11)}\\
\hline
N\_14\_52 & 2.704 & BLAGN1 & 21.10 & 1.98 & -5.04 &  -0.10 & -4.14 & 44.61 & 11 & 5 \\
N\_12\_26 & 2.471 & BLAGN1 & 20.88 & 1.96 & -5.08 &  -0.16 & -3.92 & 44.52 & 14 & 22\\
N\_13\_30 & 2.405 & BLAGN1 & 20.97 & 1.98 & -4.66 &  -0.30 & -3.88 & 44.89 & 56 & 46\\
N\_12\_35 & 2.209 & BLAGN1 & 20.74 & 2.08 & -4.84 &  -0.35 & -4.08 & 44.58 & 24 & 10\\ 
N\_59\_18 & 2.921 & BLAGN1 & 20.88 & 2.01 & -4.87 &  -0.31 & -3.93 & 44.80 & 21 & 30\\\hline
\end{tabular}
\parbox[]{170mm}{The columns are: (1) unique ID for each source; (2) redshift;
    (3) optical classification (4) logarithmic median value of the $N_{\rm H}$ parameter, in units
    of ${\rm cm^{-2}}$; (5) median value of the photon index parameters; (6) logarithmic median value
    of the normalization; (7) logarithmic median value of the reflection component normalization; (8) logarithmic
    median value of the soft scattering normalization; (9) logarithmic median value of the
    2-10~keV luminosity in units of erg~s$^{-1}$; (10) rest-frame 2-10~keV photon counts of the PN
    spectrum; (11) rest-frame 2-10~keV photon counts of the MOS spectrum. The full table is available online.}
\end{center}
\end{table*}

\subsection{\label{sec:bol_edd}Bolometric luminosity and Eddington ratio estimation
for type-1 AGN}

The bolometric luminosity $\lbol$ can be best determined by integrating the
spectral energy distribution (SED) of an AGN
\citep[e.g.][]{Elvis+al+1994,Richards+al+2006}. In this work, only the
radiation directly produced by the accretion process, i.e. the optical/UV
thermal emission from the accretion disc and the hard X-ray radiation produced
by inverse-Compton scattering of the soft disc photons by a hot corona, are
included to estimate $\lbol$, i.e. $\lbol=\ldisc+\lxray$. The infrared (IR)
radiation, which is known to be reprocessed UV radiation \citep{Antonucci+1993},
is excluded. 

The calculation of the hard X-ray radiation $\lxray$ for each source is
described in Section\,\ref{sec:xray_prop}. To estimate $\ldisc$, we assume
that the optical/UV emission are originated from a standard thin accretion disc
\citep[AD][]{Shakura+Sunyaev+1973}. In the AD model, the flux of per unit area
is 
\begin{equation}
\label{eq:ad_disc}
F=\frac{3}{8\pi}\frac{G\dot{M}\mobh}{R^3}f_c(r,a)
\end{equation} 
where $R$ is
the radius and $f_c(r,a)$ is a dimensionless factor depending on black hole
spin ($a$) and dimensionless radius $r=R/R_{\rm g}$ ($R_{\rm g}=GM/c^2$,
gravitational radius). The $f_c(r,a)$ factor is ignored since only the
radiation from relatively large radii is considered. It is clear from
Equation\,\ref{eq:ad_disc} that the full disc spectrum is determined by the BH
mass $M_{\rm BH}$ and the accretion rate $\dot{M}$. We estimate the $M_{\rm
BH}$ from the optical broad lines as discussed in Section\,\ref{subsec:mbh_edd}.
The accretion rate $\dot{M}$ is calculated from the empirical relation between
the $\dot{M}$, the $M_{\rm BH}$ and the optical monochromatic luminosity at
$4681\AA$ ($L_{\rm opt,45}$) presented in \citet[][see their equation
8]{Davis+Laor+2011}. To calculate the $L_{\rm opt,45}$, we assume a power-law
form for the optical continua, i.e.,
$f(\lambda)\propto\lambda^{-\alpha_{\lambda}}$, where the index
$\alpha_{\lambda}$ is calculated from the measured $L_{3000}$ and $L_{5100}$ if
both are available, or assuming $\alpha_{\lambda}=1.5$ if only one of them is
available or only monochromatic luminosities measured at the other wavelength
are available.

\begin{figure} 
    \begin{center}
    \includegraphics[width=\columnwidth]{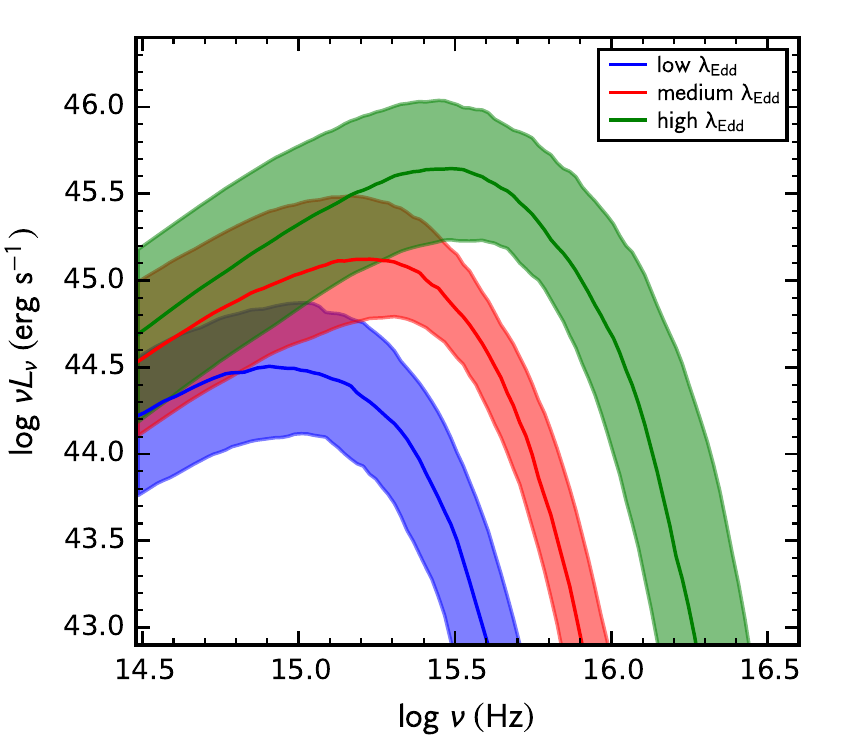}
    \caption{\label{fig:edd_bol} The median SED for the three
    different $\ledd$ samples (see Section\,\ref{sec:sub_sample}).
    The filled area marked the 25th and 75th percentile range.
    }
\end{center} 
\end{figure}

\begin{figure*} 
    \begin{center}
    \includegraphics[width=0.49\textwidth]{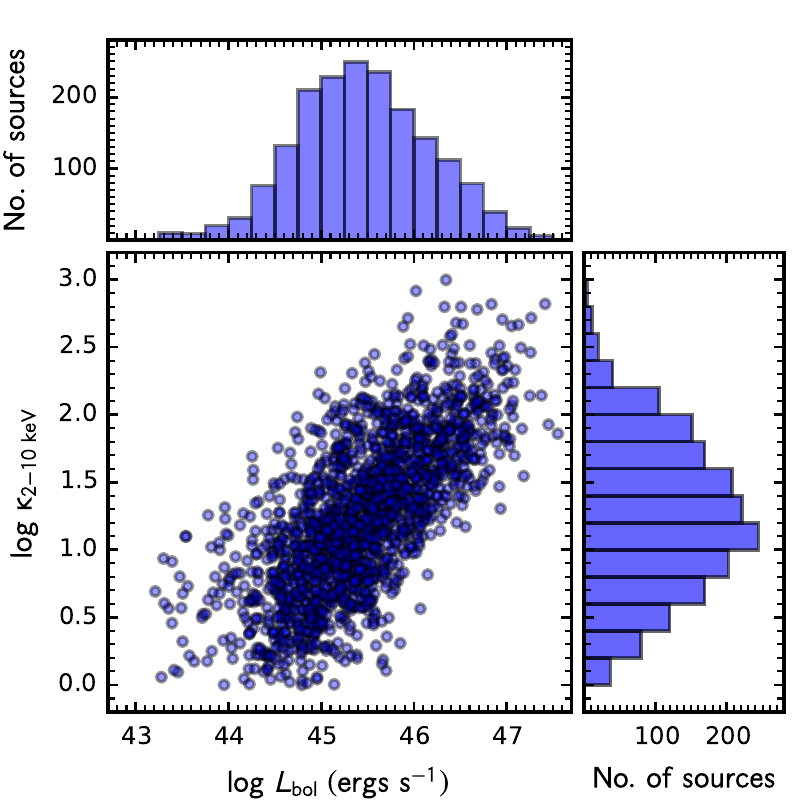}
    \includegraphics[width=0.49\textwidth]{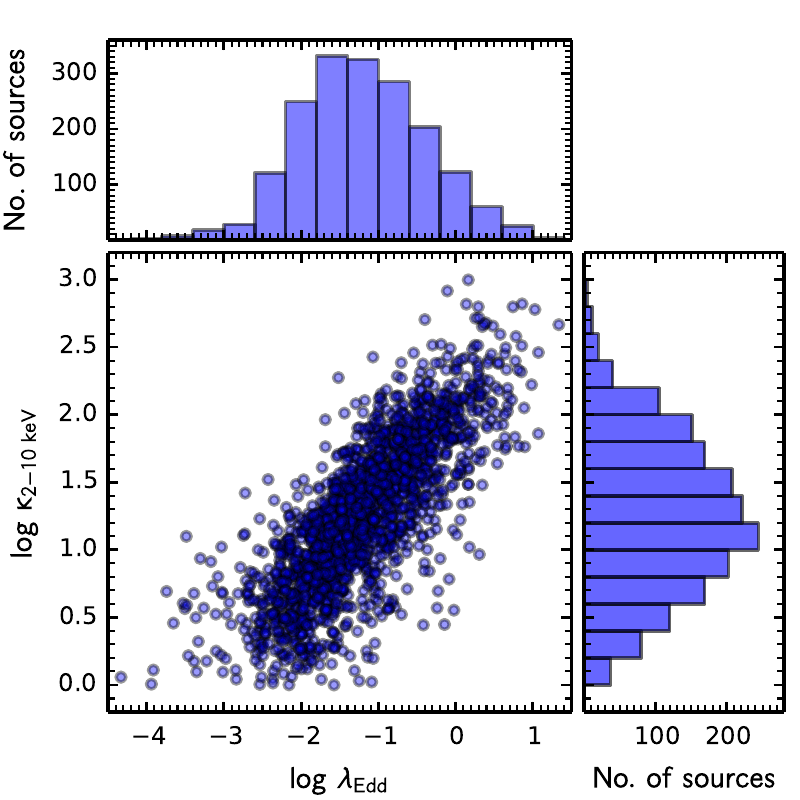}
    \caption{Left: \label{fig:bolcor_bol}The main panel shows the distribution of the
    sources in the $\bolc-\lbol$ plane (blue points). The top panel and 
    right panel show the distribution of the $\lbol$ and $\bolc$ for the sources
    in the sample, respectively. Right: the same as the left plot, but for the 
    $\bolc$ versus $\ledd$.
    }
\end{center} 
\end{figure*}

At a certain radius, the effective temperature $T_{\rm eff}$ of the disc is
$T_{\rm eff} = (F/\sigma)^{1/4}$, assuming that the disc emits locally as
blackbody.  To calculate the disc luminosity $L_{\rm disc}$, we then integrated
it over the disc surface with inner radius $R_{\rm in}=20R_{\rm g}$ and outer
radius $R_{\rm out}= 6000R_{\rm g}$ in the
$3.0\times10^{14}-4.8\times10^{17}$\,Hz frequencies range. As an illustration,
we show the median SED of the three different $\ledd$ samples (see
Section\,\ref{sec:sub_sample}) in Fig.\,\ref{fig:edd_bol}.  This shows that, on
average, the accretion disc becomes more luminous and hotter with increasing
$\ledd$.

\begin{table*}
%\begin{minipage}{120mm}
\bgroup
\setlength\tabcolsep{10pt}
\begin{center}
\caption{\label{tab:opt_cata}The optical properties of the type-1 AGN in our sample}
\begin{tabular}{|l|c|c|c|c|c|c|c|c|c|}
\hline\hline
  \multicolumn{1}{|c|}{UXID} &
  \multicolumn{1}{c|}{$z$} &
  \multicolumn{1}{c|}{$\log\mobh$} &
  \multicolumn{1}{c|}{$\log{L_{1350}}$} &
  \multicolumn{1}{c|}{$\log{L_{1700}}$} &
  \multicolumn{1}{c|}{$\log{L_{3000}}$} &
  \multicolumn{1}{c|}{$\log{L_{5100}}$} &
  \multicolumn{1}{c|}{$\log\ldisc$} &
  \multicolumn{1}{c|}{$\log\lbol$} &
  \multicolumn{1}{c|}{$\log\ledd$} \\
  \multicolumn{1}{|c|}{(1)}&
  \multicolumn{1}{c|}{(2)} &
  \multicolumn{1}{c|}{(3)} &
  \multicolumn{1}{c|}{(4)} &
  \multicolumn{1}{c|}{(5)} &
  \multicolumn{1}{c|}{(6)} &
  \multicolumn{1}{c|}{(7)} &
  \multicolumn{1}{c|}{(8)} &
  \multicolumn{1}{c|}{(9)} &
  \multicolumn{1}{c|}{(10)} \\
\hline
N\_0\_16 & 1.597  &  8.10  & ---    & 44.85 &  44.84 &  ---   & 45.81 & 45.83 & -0.39\\
N\_0\_27 & 1.741  &  8.82  & 44.60  & 44.66 &  44.97 &  ---   & 45.31 & 45.36 & -1.58\\
N\_0\_29 & 0.996  &  7.73  & ---    & ---   &  44.32 &  44.47 & 45.70 & 45.70 & -0.15\\
N\_0\_30 & 1.687  &  9.68  & 45.70  & 45.64 &  45.57 &  ---   & 45.31 & 45.49 & -2.30\\
N\_0\_32 & 0.985  &  8.11  & ---    & ---   &  44.13 &  44.34 & 45.14 & 45.16 & -1.06\\\hline
\end{tabular}
\parbox[]{160mm}{The columns are: (1) unique ID for each source; (2) redshift;
    (3) $\mobh$ estimate using method described in Section\,\ref{subsec:mbh_edd};
    (4)-(7) optical monochromatic luminosity in units of erg\,s$^{-1}$; (8) the disc luminosity is
    calculated with the method presented in Section\,\ref{sec:bol_edd};
    (9) the bolometric luminosity in units of erg~s$^{-1}$; (10) $\ledd$. The full table is available
        online.}
\end{center}
\egroup
\end{table*}

The bolometric luminosity is then computed as the sum of the disc luminosity
and the $2-10$\,keV X-ray luminosity. The $\ledd$, given by $\ledd= L_{\rm
bol}/L_{\rm Edd}$, as a function of the $\mobh$ is plotted on the
central panel of Fig.\,\ref{fig:mbh_edd}, while the $\ledd$ distribution is
shown in the horizontal histogram. Fig.\,\ref{fig:bolcor_bol} shows the
$2-10~\mathrm{keV}$ X-ray bolometric correction ($\bolc=\lbol/\lxray$) versus
$\lbol$ (left panel) and the $\ledd$ (right panel). Tight correlations
between the $\bolc$ and $\lbol$ (Pearson correlation coefficient $r=0.71$)
as well as the $\ledd$ (Pearson correlation coefficient $r=0.78$) are found in
our sample as clearly shown in Fig.\,\ref{fig:bolcor_bol}. A similar trend,
that the sources with higher bolometric luminosities have higher bolometric
corrections, was found in \citet{Lusso+al+2012}. A relation between the $\bolc$
and $\ledd$ is also found in several previous works
\citep{Vasudevan+Fabian+2007, Vasudevan+Fabian+2009, Vasudevan+al+2009,
Lusso+al+2012}. The tight $\bolc-\lbol$ relation shown in
Fig.\,\ref{fig:bolcor_bol} as well as the correlation between $\bolc$ and
$\ledd$, rather than intrinsic, can however be a natural outcome of our
flux-limited sample, which will lead to a correlation between the monochromatic
optical luminosity $L_\mathrm{mon,opt}$ and the X-ray luminosity $\lxray$. We
tested the significance of the correlation by means of simulations. We
generated a sample which the $\log{N}-\log{S}$ of both the X-ray and optical
flux follows a power-law form, and has a redshift distribution similar to our
AGN sample. We then applied the flux cut-off,
$f_\mathrm{0.5-10\,keV}>1\times10^{-15}\,\mathrm{erg\,s^{-1}\,cm^{-2}}$ and
$R_{4681}<22\,\mathrm{mag}$, on the simulated sources. Using the
$L_\mathrm{mon,opt}$ and $\lxray$ of the simulated data we obtain a Pearson
correlation coefficient value $r$ of $\sim0.57$, which is {\it smaller} than the
correlation coefficient found in the observed data ($r\sim0.71$). Thus the
observed correlations in our sample should not be purely due to the selection
bias. This result is also consistent with the conclusion that the strong correlation
between the monochromatic UV luminosity at 2500{\AA} and the X-ray luminosity at
2\,keV is independent of redshift \citep{Risaliti+Lusso+2015}. The
distributions of the $\lbol$ and $\bolc$ are shown in
Fig.\,\ref{fig:bolcor_bol} (top panel: $\lbol$, right panel: $\bolc$). The
$\bolc$ found in our sample, with a mean value of $\sim18$, are consistent with
the values found in previous works \citep[e.g.][]{Vasudevan+Fabian+2007,
Lusso+al+2012}.

The optical spectral properties of the type-1 AGN as well as the estimated
$\mobh$, $\lbol$ and $\ledd$ are included in Table\,\ref{tab:opt_cata} (the
full table is available online). The optical monochromatic luminosities are
estimated via spectral fitting procedure described in
Section\,\ref{subsec:mbh_edd}.

%%*************************************************
%%************Section: Sample selection*************
%%*************************************************
\section{X-ray spectral stacking of the type-1 AGN}\label{sec:xraystack}

As shown in Fig.\,\ref{fig:pho_dist}, the majority of the sources in our
sample have low photon counts. Spectral stacking is an effective way to
obtain a composite spectrum with high S/N in this case. Here we adopted the
rest-frame X-ray spectral stacking method to investigate the average properties of AGN.

We studied various type-1 AGN sub-samples selected on the basis of their
physical properties, such as the X-ray luminosity, redshift, $\ledd$ as well as
$\mobh$. For each source the PN or MOS spectra is used if a) the rest-frame
2-10\,keV photon counts is larger than 10 and less than 200; b) the fitted MOS/PN
normalization factor is in the range 0.5-1.5 so that we can reject poor
quality spectra and sources for which one of the spectra fall, for example, at
the edge or in the gap between CCDs. About 10 per cent of the total type-1
sources are discarded after applying these criteria. The rest of the sources
form the main sample for the X-ray spectral stacking. The median redshift and
$\lxray$ of this main sample are 0.87 and $3.1\times10^{44}$\,erg\,s$^{-1}$,
respectively.

%%*************************************************
%%*********subsection: ****************************
%%*************************************************
\subsection{\label{sec:sub_sample}Sub-samples}

We first define four redshift-luminosity sub-samples. The sub-samples are
selected in a way that the stacked spectra have similar S/N. The main sample
is divided into two redshift bins, the low-redshift bin including sources with
redshift $z<1.36$ and the high-redshift bin with $z>1.36$.  Each redshift bin
is further split into two X-ray luminosity bins: low
($42.00<\log\lxray<44.00~{\rm erg~s^{-1}}$) and medium
($44.00<\log\lxray<44.66~{\rm erg~s^{-1}}$) luminosity subsets for the low
redshift bin, medium ($44.00<\log\lxray<44.66~{\rm erg~s^{-1}}$) and high
($44.66<\log\lxray<46.00~{\rm erg~s^{-1}}$) luminosity subsets for the high
redshift bin. Thus, in total, we have four luminosity-redshift sub-samples,
         namely: low-redshift low-luminosity (LL), low-redshift medium
luminosity (LM), high-redshift medium luminosity (HM) and high-redshift high
luminosity (HH) sub-samples. The details of each sub-sample are shown in
Table\,\ref{tab:sub_samples}, and illustrated in Fig\,\ref{fig:lum_red_dist}.

We also define three $\mobh$ sub-samples, namely: low $\mobh$ sample with
$\log\mobh<8.4$; medium $\mobh$ with $8.4<\log\mobh<8.9$; high $\mobh$ with
$\log\mobh>8.9$, and three $\ledd$ sub-samples: low $\ledd$ sample with
$\log\ledd<-1.62$; medium $\ledd$ with $-1.62<\log\ledd<-0.95$; high $\ledd$
with $\log\ledd>-0.95$. Details of the $\mobh$ and $\ledd$ sub-samples
can be found in Table\,\ref{tab:sub_samples}.

\begin{table}
%\begin{minipage}{120mm}
\bgroup
\setlength\tabcolsep{4.8pt}
\begin{center}
\caption{\label{tab:sub_samples}Properties of the type-1 AGN sub-samples}
\begin{tabular}{lccccccc}\\\hline\hline
Sample	& $\tilde{z}$ & $\tilde{L}_{\rm{2-10~keV}}$ & $\tilde{M}_{\rm{BH}}$ & $\tilde{\lambda}_{\rm{Edd}}$ & \#src & \#spec & cts \\
         & (1)   & (2)    & (3)   & (4)    & (5) &  (6)  &  (7)  \\\hline
LL       & 0.77  & 43.67  & 8.22  &  -1.41 & 418 & 708   & 25581\\
LM       & 1.14  & 44.21  & 8.57  &  -1.50 & 270 & 500   & 25367\\
HM       & 1.82  & 44.39  & 8.71  &  -1.20 & 517 & 858   & 24660\\
HH       & 2.25  & 44.82  & 8.97  &  -0.87 & 256 & 461   & 24717\\
Low Edd  & 1.24  & 44.13  & 8.81  &  -2.00 & 481 & 826   & 34191\\
Med Edd  & 1.44  & 44.28  & 8.63  &  -1.29 & 496 & 863   & 34985\\
Hig Edd  & 1.67  & 44.33  & 8.25  &  -0.44 & 557 & 949   & 35271\\
Low MBH  & 1.11  & 43.98  & 8.07  &  -0.74 & 576 & 973   & 36008\\
Med MBH  & 1.42  & 44.24  & 8.64  &  -1.41 & 548 & 936   & 36320\\
Hig MBH  & 1.77  & 44.54  & 9.12  &  -1.59 & 410 & 729   & 32119\\\hline
\end{tabular}
\parbox[]{84mm}{The columns are: (1) median redshift; (2) median
    2-10~keV luminosity in unit erg~s$^{-1}$; (3) median $\mobh$;
    (4) median $\ledd$; (5) number of sources; (6) number
    of spectra, PN and MOS spectra are counted independently; (7) total 2-10~keV net photon counts}
\end{center}
\egroup
%\end{minipage}
\end{table}

%%*************************************************
%%*************Section: Data analysis**************
%%*************************************************
\subsection{\label{sec:spec_stack}Spectral stacking method}

The rest-frame stacking method adopted in this work is similar to that presented in
\citet[][see also \citealt{Falocco+al+2012}]{Corral+al+2008}. Here we outline the main procedures of this method.
\begin{enumerate} 
\item Unfolding the spectrum: The intrinsic source spectra can be approximately
	reconstructed with the instrumental effects eliminated by unfolding the
	observed spectra. A model-dependent unfolding method is used to unfold the
	observed spectra. The {\tt eufspec} command in {\sc Xspec}
	\citep{Arnaud+1996} is applied to calculate the unfolded spectra,
	assuming a absorbed power-law with Galactic column density fixed at
	$N_{\rm{H}}=2.0\times10^{20}\,\rm{cm}^{-2}$ given by \citet{Kalberla+al+2005}
	in the XXL region and photon index obtained from the fitted
	value using the BXA method (see Section\,\ref{sec:xray_prop}).
\item Rebin: We then correct the Galactic absorption and de-redshift the
    unfolded spectra to the source rest-frame. The de-redshifted rest-frame
    spectra have different energy bins, and thus have to be rebinned into
    unified bins before stacking. To construct a new, unified bin scale, we
    group the rest-frame 2.1-10.1~keV energy range spectrum of each source with
    a bin width of 200~eV.
\item Stacking: We obtain the stacked spectrum by averaging the rebinned
    rest-frame unfolded spectra with simple arithmetic mean. We didn't rescale
    each spectrum by the its flux due to the large uncertainty in choosing the
    proper rescaling factor for faint sources, which are the majority of our
    sample, as also pointed out in \citet{Iwasawa+al+2012}.
\end{enumerate}

The PN and MOS spectra are stacked separately, to avoid any
features induced by the stacking method and by the uncalibrated systematic
differences between MOS and PN spectra.

%%*************************************************
%%***************Subsection: Results******************
%%*************************************************
\subsection{\label{subsec:results}Results}

\begin{figure*}
    \begin{center}
    \includegraphics[width=\linewidth]{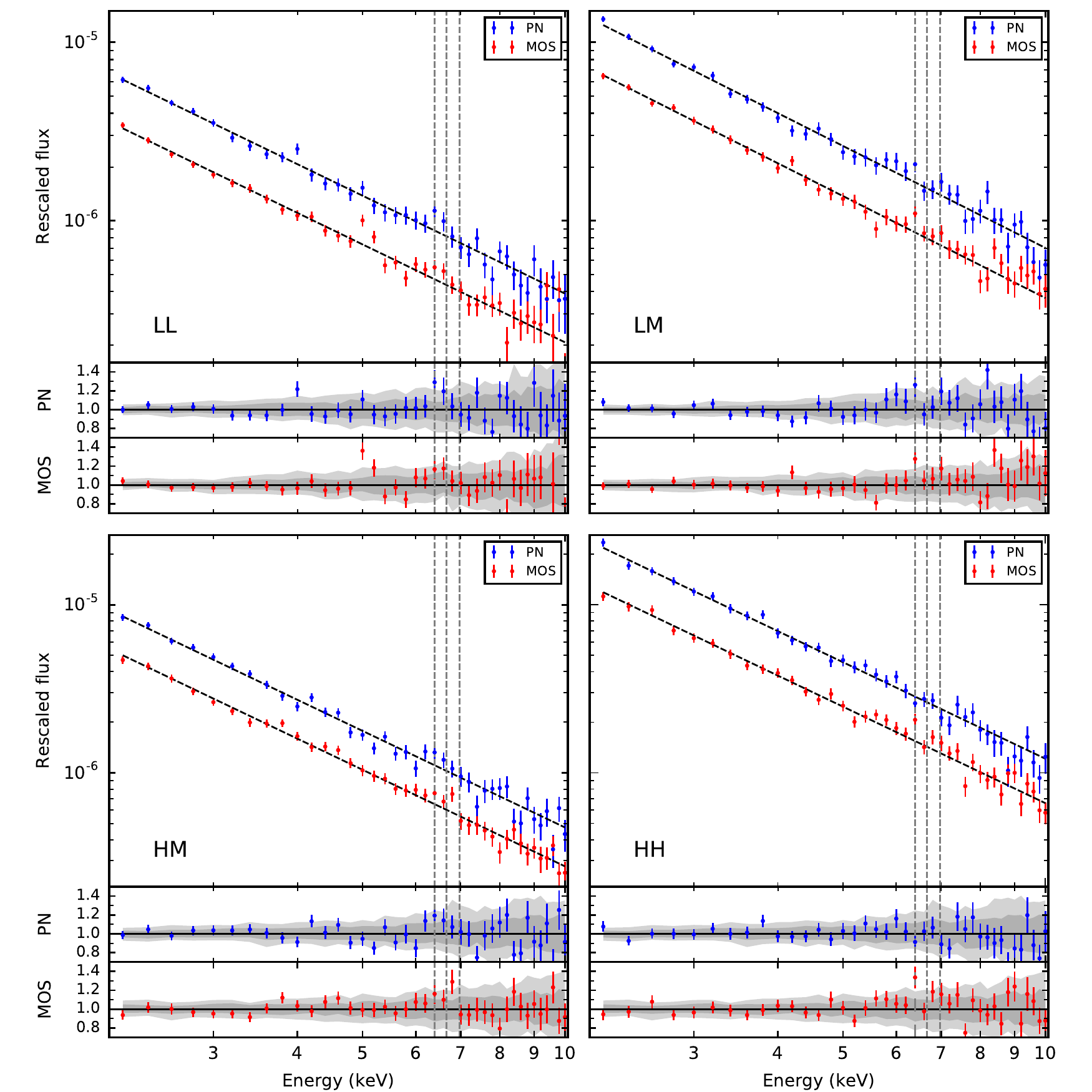}
    \caption{\label{fig:red_lum_subs}The stacked spectra of the PN (blue) and MOS (red)
    cameras for the redshift-luminosity sub-samples, with the best-fitting
    power-law continua (black dashed line) are shown in the {\it upper} plot of
    each panel (for sake of clarity, we rescaled the PN and MOS stacked
    spectra). The ratios of the stacked spectra to the best-fitting power-law
    continua are shown in the middle (PN) and bottom (MOS) plots of each
    panels. The $1\sigma$ and $2\sigma$ confidence intervals are marked as dark
    and light shaded areas, respectively. The grey dashed vertical lines are
    indications of the 6.4\,keV, 6.67\,keV and 6.97\,keV emission lines.
    Details of each sub-sample can be found in Table\,\ref{tab:sub_samples}}
\end{center} 
\end{figure*}

\begin{figure*} 
    \begin{center}
    \includegraphics[width=\linewidth]{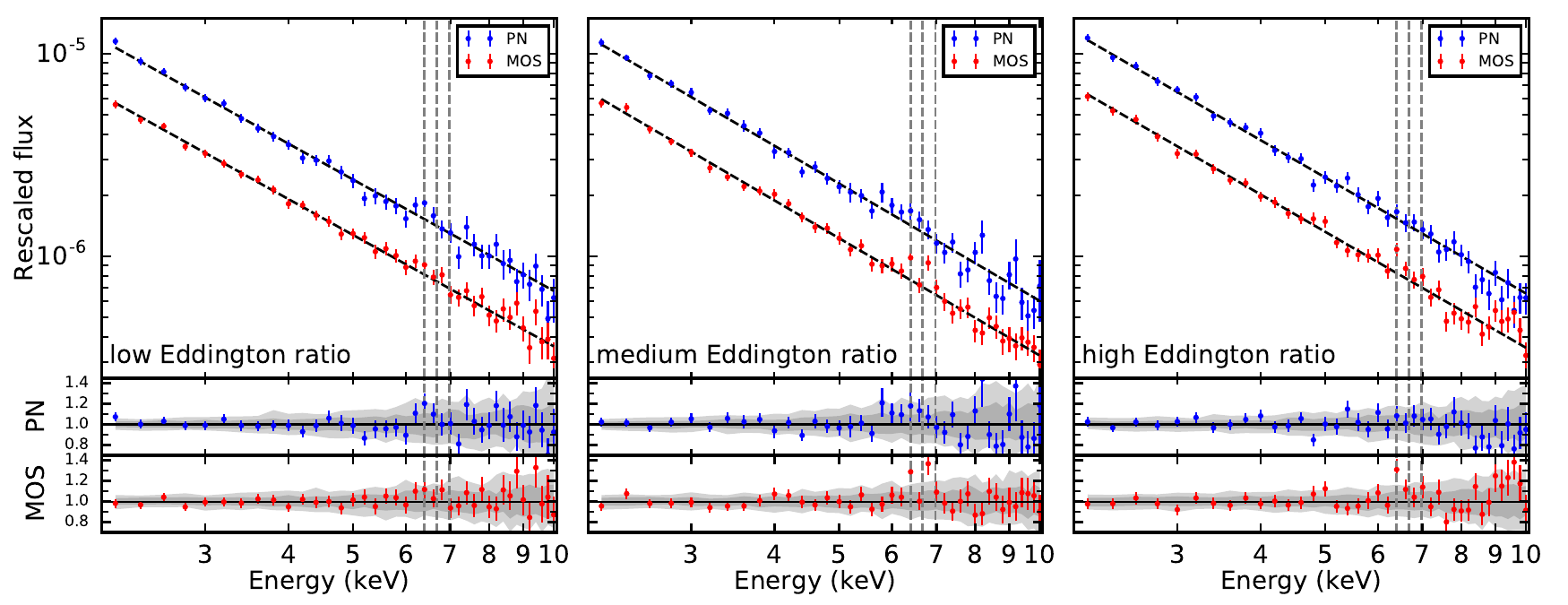}
    \includegraphics[width=\linewidth]{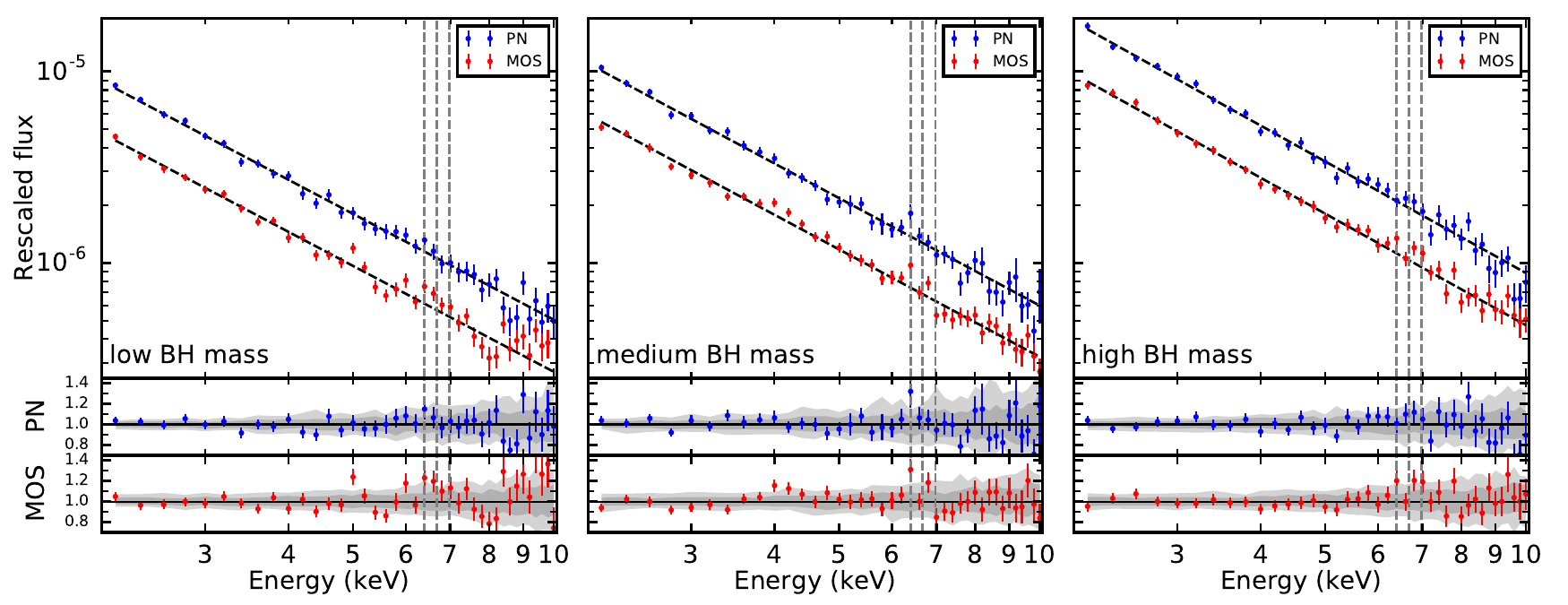}
    \caption{\label{fig:edd_mbh_subs} Same as Fig.\,\ref{fig:red_lum_subs}, but for $\ledd$
    (upper panel) and $\mobh$ (lower panel) sub-samples. Details of each sub-sample
	can be found in Table\,\ref{tab:sub_samples}}
\end{center} 
\end{figure*}

The rest-frame 2-10\,keV stacked spectra of the sub-subsamples are shown
in Figs.\,\ref{fig:red_lum_subs} (the redshift-luminosity sub-samples) and
\ref{fig:edd_mbh_subs} (upper-panel: the $\ledd$ sub-samples; lower-panel: the
$\mobh$ sub-samples). The PN and MOS stacked spectra are marked with blue and
red points, respectively. In each plot, the ratio of the data to a
best-fitting power-law continuum is also presented. There is evidence for
excess emission at around 6.4\,keV above the power-law continuum which could be
attributed to the Fe neutral K$\alpha$ emission line. In addition, potential
highly ionized Fe K emission lines (e.g. the 6.67\,keV Fe\,{\sc xxv} and
6.97\,keV Fe\,{\sc xxvi}) are also detected in some of the stacked spectra,
e.g. the LM sub-sample. No significant absorption feature is found in the
stacked spectra of the sub-samples, which is consistent with the optical
classification (the fraction of optical type-1 sources with $N_{\rm
H}>10^{21.5}{\rm cm^{-2}}$ is quite low in each sub-sample, thus they will not
dominate the stacked spectra).

Except for the Fe K emission lines, unidentified line features, especially with
line energy peaked at the low energy band (e.g. $<5.5$\,keV), are also shown in
some of the stacked spectra. These line features can be due to systematical
uncertainties of the data, as in most cases they are present only in one of the
PN and MOS spectra. We carry out simulations to test whether these lines are
artefacts and also to estimate the significance of the Fe K emission
line. For each source in our sample, assuming a pure power-law model, we
simulate 100 faked spectra using {\sc Xspec}. To match the observations, the
exposure time, auxiliary and response matrix files for the sources in the
sample are used in the simulations. For each sub-sample, we generate 100
stacked spectra by applying the same stacking method to the simulated source
spectra. We further estimate the $1\sigma$ and  $2\sigma$ confidence intervals
by calculating the 3rd, 16th, 84th, 97th percentile values in each energy bin,
which are shown as grey shadowed area in Fig\,\ref{fig:red_lum_subs} and
\ref{fig:edd_mbh_subs}. Contrary to what has been found in previous works
\citep[e.g.][]{Corral+al+2008}, there are spurious lines features shown in the
stacked spectra. This can be explained by the fact that our data on average
have much lower S/N, thus larger systematic uncertainties, compared to the
sample used in previous works.  These lines, which are significantly detected
in only one of the PN and MOS simulated stacked spectra, are not ubiquitously
found in the stacked spectra. A spurious emission line feature at 6.4~keV
($\sim$\,6.7\,keV or $\sim$\,6.97\,keV), detected in both the PN and MOS
stacked spectra, is found in about 6 (9) out of the 100 simulated spectra. The
EWs of the spurious emission lines at 6.4\,keV are around $\sim50\pm20$\,eV.

The energy resolution of both EPIC PN and MOS cameras is $\approx0.15\,\mathrm{keV}$\,(FWHM) at 6.4\,keV.
A narrow line can be broadened by the stacking method described in Section\,\ref{sec:spec_stack}, especially
in the unfolding procedure, due to instrumental effects. To examine this, we generate a series of spectra with high S/N using
an intrinsically unresolved Gaussian line model (line width $\sigma=1\,\mathrm{eV}$).
These simulated spectra have the same normalization, but different redshift, $z=0,1,2$.
We find that an intrinsically unresolved line manifests itself as a Gaussian profile in the
unfolded spectrum. The line width increases with redshift and is in the range of
$75-180\,\mathrm{eV}$. These results are consistent with the conclusions reached by previous
work \citep[e.g.][]{Corral+al+2008, Iwasawa+al+2012, Falocco+al+2012, Liu+al+2015}.

We use {\sc Xspec} (version 12.8) to fit the stacked spectra. The {\tt FTOOLS}
task {\tt flx2xsp} is used to convert the flux spectra into fits format which
can be fitted using {\sc xspec}. A base-line model consisting of a power-law
and a Gaussian components, with all the parameters free, is used to jointly fit
the PN and MOS stacked spectra.  In some cases, the central energy of the
Gaussian is fixed at 6.4\,keV while the line width is fixed at 120\,eV
\footnote{As mentioned above, an unresolved narrow line will be broadened due to instrumental
effects, and the artificial line broadening is redshift dependent, i.e. the line will be
broader for high-redshift sources. We thus tried to fix the line width at
different values (e.g. 100-150\,eV). However, we did not
find any significant difference. All calculated EWs of the emission line
are consistent within their mutual uncertainties and will not affect our
conclusion.}, if these parameters cannot be well constrained due to the low S/N
in the Fe line region of the stacked spectra.

\subsubsection{The redshift-luminosity sub-samples}

A significant emission line feature peaking at 6.4~keV, which must be the
neutral Fe K$\alpha$ line, is detected ($>2\sigma$ level) in the stacked
spectra of both the LL and LM sub-samples, as shown in the upper panel of
Fig.\,\ref{fig:red_lum_subs}, although it is only marginally detected at the
$2\sigma$ level in the MOS stacked spectrum of the LL sub-sample. The base-line
model can fit the stacked spectra of the LL sub-sample well with the
best-fitting photon index $\Gamma=1.85\pm0.03$. The EW of the line is
$125\pm49$\,eV, which is significantly higher than the EWs of the spurious
6.4\,keV lines found in the simulations. In addition to the neutral Fe
K$\alpha$ line, an emission line feature peaked at around 7.0\,keV is barely
detected at $2\sigma$ level in both the PN and MOS stacked spectra of the LM
sub-sample. Thus a Gaussian component is added to the base-line model. The line
width of the $\sim7.0$\,keV emission line, which can not be well constrained,
is fixed at 120\,eV. We also fixed the line width of the 6.4\,keV emission line
at its best-fitting value, i.e. $\sigma=88$\,eV.  The best-fitting photon index
is $1.93\pm0.02$. The line energy of the two Gaussian components are
$6.37\pm0.06$ and $7.0\pm0.08$\,keV (possibly corresponding to the highly
ionized 6.97\,keV Fe\,{\sc xxvi} emission line), respectively. The neutral and
highly ionized emission lines have comparable EWs, i.e. $\mathrm{EW}_{6.4\,{\rm
keV}}=82\pm25$ and $\mathrm{EW}_{6.97\,{\rm keV}}=72\pm33\,{\rm eV}$.

The neutral Fe K$\alpha$ line is only marginally detected at $\sim2\sigma$
level in the stacked spectra of the HM sub-sample (bottom-left panel of
Fig.\,\ref{fig:red_lum_subs}). A highly ionized Fe K emission line feature is
present only in the MOS stacked spectrum of the HM sub-sample. It can be either
due to the systematic uncertainty or a true high ionization emission line. In
the first case, we fit the stacked spectra, excluding the $6.8\,{\rm keV}$ data
point in the MOS spectrum, with the base-line model. The best-fitting slope of
the power-law continuum is $1.93\pm0.02$. The line energy of the Gaussian
component is $6.43\pm0.10\,{\rm keV}$ with the line width
$\sigma=188^{+104}_{-71}\,{\rm eV}$. The EW of the emission line is
$99\pm46\,{\rm eV}$. In the second case, the emission line detected in the PN
spectrum should be a mixture of a neutral emission line with a high ionization
emission line. We fit the PN and MOS spectra with the base-line plus a Gaussian
component model. The line widths of the two Gaussian components are fixed at
120\,eV. The best-fitting photon index is $1.93\pm0.02$. The line energies of
the neutral and high ionization emission line are $6.34\pm0.08$ and
$6.73\pm0.08\,{\rm keV}$, respectively. The equivalent width of the neutral
emission line is $66\pm25$\,eV while the highly ionized emission line has a
line EW of $69\pm34$\,eV. The EW of the neutral emission line is consistent
with the first case.

No significant line features are detected in the PN stacked spectrum of the HH
sub-sample (bottom-right panel of Fig.\,\ref{fig:red_lum_subs}), while an
emission line peaking at $\sim6.4\,{\rm keV}$ and a possible highly ionized
emission line are present in the MOS stacked spectrum. We fixed the line energy
of the Gaussian component at 6.4\,keV and the line width at 120\,eV. The
best-fitting photon index is $1.92\pm0.02$ and the EW of the emission line is
less than 60\,eV.  We also try to only fit the MOS spectrum with the base-line
model plus a Gaussian component. The line energy of the first emission line,
which is consistent with an neutral Fe K$\alpha$ line as shown in the
bottom-right panel of Fig.\,\ref{fig:red_lum_subs}, is fixed at 6.4\,keV.  The
best-fitting photon index is $1.89\pm0.03$. The line width of the 6.4\,keV line
is found to be $\lesssim 100\,{\rm eV}$. The EW of the neutral Fe K$\alpha$
line is $68\pm40$\,eV. The line energy of the second Gaussian component is
$6.90\pm0.10$\,keV (to estimate the uncertainty, the line widths of the two
Gaussian components are fixed to their best-fitting values) with EW of
$80\pm44$\,eV. The line width of this emission line is found to be $121\,{\rm
eV}$ with an upper limit of 303\,eV.

\begin{table*}
\setlength\tabcolsep{0.7mm}
\begin{center}
\caption{\label{tab:fit_subsamples}Estimated parameters for each sub-sample}
\begin{tabular}{@{\extracolsep{8pt}}lcccccccccccc@{}}\\\hline\hline
    & \multicolumn{4}{c}{JOINT} &  \multicolumn{4}{c}{EPIC PN} & \multicolumn{4}{c}{EPIC MOS} \\\cline{2-5}\cline{6-9}\cline{10-13}\\[-4.2mm]
\raisebox{2.5mm}{Samples} & $\Gamma^{a}$ & $E_{\rm Gau}$(keV) & $\sigma$(eV) & EW(eV)  & $\Gamma^{b}$ & $E_{\rm Gau}$(keV) & $\sigma$(eV) & EW(eV)  & $\Gamma^{b}$ & $E_{\rm Gau}$(keV) & $\sigma$(eV) & EW(eV) \\[-0.8mm]\hline
LOW LOW        & $1.85$ & $6.46\pm0.08$ & $190^{+ 91}_{- 86}$ & $125\pm49$ & $1.87$ & $6.46\pm0.08$ & $149^{+129}_{-147}$ & $139\pm64$ & $1.83$ & $6.46\pm0.13$ & $216^{+119}_{-203}$ & $110\pm 67$\\[0.5mm]
LOW MED        & $1.93$ & $6.37\pm0.06$ & $ 88^{+ 65}_{-86.}$ & $ 82\pm25$ & $1.97$ & $6.33\pm0.08$ & $120^{f   }       $ & $111\pm48$ & $1.89$ & $6.40\pm0.07$ & $ 83^{+ 83}_{- 83}$ & $ 73\pm 42$\\[0.5mm]
HIG MED$^*$   & $1.93$ & $6.43\pm0.10$ & $188^{+104}_{- 71}$ & $ 99\pm46$ & $1.97$ & $6.48\pm0.13$ & $199^{+128}_{- 78}$ & $136\pm72$ & $1.89$ & $6.43\pm0.14$ & $120^{f   }       $ & $ 59\pm 32$\\[0.5mm]
HIG HIG        & $1.91$ & $6.40^{f}   $ & $120^{f   }       $ & $ <60    $ & $1.95$ & $6.40^{f}   $ & $120^{f   }       $ & $ <30    $ & $1.89$ & $6.40^{f}   $ & $ <100            $ & $ 68\pm 40$\\[0.0mm]\hline
LOW EDD        & $1.84$ & $6.38\pm0.10$ & $150^{+103}_{-142}$ & $ 73\pm30$ & $1.89$ & $6.41\pm0.09$ & $160^{+ 93}_{- 65}$ & $111\pm50$ & $1.80$ & $6.40^{f}   $ & $120^{f   }       $ & $ <70     $\\[0.5mm]
MED EDD$^*$    & $1.95$ & $6.38\pm0.24$ & $120^{+260}_{- 50}$ & $ 88\pm30$ & $1.97$ & $6.42\pm0.14$ & $120^{f   }       $ & $ 98\pm46$ & $1.92$ & $6.37\pm0.08$ & $ 63^{+ 61}_{- 63}$ & $ 70\pm 59$\\[0.5mm]
HIG EDD        & $1.91$ & $6.46\pm0.06$ & $120^{f   }       $ & $ 60\pm26$ & $1.94$ & $6.40^{f}   $ & $120^{f   }       $ & $ <56    $ & $1.87$ & $6.40^{f}   $ & $ 71^{+ 52}_{- 71}$ & $ 71\pm 32$\\[0.0mm]\hline
LOW MBH        & $1.86$ & $6.48\pm0.14$ & $ <470            $ & $ 73\pm30$ & $1.88$ & $6.43\pm0.20$ & $123^{+ 77}_{-123}$ & $ 63\pm40$ & $1.83$ & $6.46\pm0.09$ & $120^{f   }       $ & $ 67\pm 47$\\[0.5mm]
MED MBH$^*$    & $1.88$ & $6.40^{f}   $ & $ <104            $ & $ 72\pm30$ & $1.93$ & $6.40\pm0.07$ & $108^{+127}_{- 96}$ & $113\pm47$ & $1.84$ & $6.40^{f}   $ & $ 69^{+ 46}_{- 69}$ & $ 71\pm 31$\\[0.5mm]
HIG MBH        & $1.96$ & $6.32\pm0.11$ & $119^{+ 69}_{- 54}$ & $ 45\pm20$ & $1.96$ & $6.40^{f}   $ & $120^{f   }       $ & $ <54    $ & $1.94$ & $6.36\pm0.07$ & $120^{f   }       $ & $ 64\pm 32$\\\hline
\end{tabular}

\parbox[]{20.0cm}{$^{*}$:Only the results fitted with the base-line plus a Gaussian component model are shown here.

    $a$: The typical $1\sigma$ uncertainty is 0.02.
    
    $b$: The typical $1\sigma$ uncertainty is 0.03.

    $f$: The line width of the Gaussian component is fixed.
}
\end{center}
\end{table*}

\subsubsection{The Eddington ratio sub-samples}

As shown in Fig. \ref{fig:edd_mbh_subs}, an emission line feature peaking at around 6.4\,keV is marginally seen in the
stacked PN spectrum ($\sim2\sigma$ level) of the low $\ledd$ sub-sample, while
it is only tentatively detected in the MOS stacked spectrum ($\gtrapprox1\sigma$
level). The best-fitting photon index of the power-law continuum in the
base-line model is $1.84\pm0.02$. The EW of the Gaussian component, with
$E=6.38^{+0.10}_{-0.08}$\,keV and $\sigma=150^{+104}_{-150}$\,eV, is
$73\pm30$\,eV. 

Two emission line features are detected at $2\sigma$ level in the MOS stacked
spectrum of the medium $\ledd$ sub-sample, while only a broad line feature
(e.g. $\sigma\sim400\,{\rm eV}$) is found in the PN stacked spectrum. We first
jointly fit the stacked spectra with the base-line model plus a Gaussian
component. The line energy of the highly ionized emission line is fixed at its
best-fitting value, $E=6.86$\,keV. The best-fitting photon index of the
power-law component in the base-line model is $\Gamma=1.95\pm0.02$. The line
energy of the neutral emission line is $6.38\pm0.24$\,keV with the line width
$\sigma=120^{+260}_{-50}$\,eV. The EW of the neutral emission line is
$88\pm30$\,eV. The EW of the high ionization emission line is $56\pm21$\,eV,
with an upper limit of the line width $\sigma=101$\,eV. We also fit the stacked
spectra with only the base-line model, excluding the energy range around
$\sim6.8$\,keV in the MOS stacked spectrum. In this case, the best-fitting
slope of the continuum as well as the parameters of the Gaussian component in
the base-line model are in good agreement with the results from the base-line
model plus a Gaussian component.

Although an emission line at 6.4\,keV is not significantly detected in the PN
stacked spectrum of the high $\ledd$ sub-sample, there is evidence for such an
emission feature in the MOS stacked spectrum ($>2\sigma$ level). We fixed the
line width at 120~eV. The best-fitting photon index is $1.91\pm0.02$ if the
stacked spectra are fit with the base-line model. The line energy is
$E=6.46\pm0.06$, consistent with the neutral Fe K$\alpha$ line. The EW of the
line is $60\pm26$\,eV.

\subsubsection{The BH mass sub-samples}

The stacked spectra of the low $\mobh$ sub-sample are fitted with the base-line
model. The best-fitting photon index of the continuum is $1.86\pm0.03$. The
line energy of the Fe K line, which is detected barely at $\sim2\sigma$ level,
is $6.48\pm0.14\,{\rm keV}$ with line width $\sigma=99\,{\rm eV}$ (with an
upper limit of $\sim470\,{\rm eV}$). The EW of the line is $73\pm30\,{\rm eV}$.

An emission peaked at $\sim6.4\,{\rm keV}$ is significantly detected
($>2\sigma$ level) in both the PN and MOS stacked spectra of the medium $\mobh$
sub-sample. A line feature with energy at $\sim6.7\,{\rm keV}$ is also
presented in the MOS stacked spectrum. We first fit the stacked spectra with
the base-line model, excluding the line feature shown in the MOS stacked
spectrum.  The slope of the power-law continuum is $1.88\pm0.02$. The line
energy of the Gaussian component is consistent with the neutral Fe K$\alpha$
line, $E=6.38\pm0.04$\,keV. The best-fitting line width is 86\,eV, with an
upper limit of 121\,eV. The EW of the neutral line is $89\pm28\,{\rm eV}$. We
also fit the stacked spectra with the base-line model plus a Gaussian
component. The line energy of the two Gaussian components are fixed at 6.4 and
6.7\,keV, respectively.  The line width of the neutral emission line is left to
be free, while it is fixed at 120\,eV for the 6.7\,keV emission line. The
fitting results are consistent with the results in the first case, i.e.
$\Gamma=1.88\pm0.02$, $\sigma<104\,{\rm eV}$ and ${\rm EW}=72\pm30\,{\rm eV}$.
The EW of the high ionized emission line is 24\,eV with an upper limit of
50\,eV.

A potential highly ionized emission line is shown in both the PN and MOS
stacked spectra of the high $\mobh$ sub-sample, while a neutral Fe K emission
line is only presented in the MOS stacked spectrum ($\sim2\sigma$ level). We
fit the stacked spectra with the base-line model plus a Gaussian component. The
line width of the neutral Fe K emission line is fixed at 120\,eV while the line
energy of the highly ionized emission line is fixed at its best-fitting value
$E=6.85$\,keV. The best-fitting photon index is $1.96\pm0.02$. The line energy
of the neutral emission line is $6.32\pm0.11$\,keV.  The line width of the high
emission line can be well constrained, i.e.  $\sigma=119^{+69}_{-54}$\,eV. The
EWs of the neutral and ionized emission line are $45\pm20$ and $64\pm30$\,eV,
respectively.

The results of the spectral fitting can be found in
Table\,\ref{tab:fit_subsamples}. We also fit the PN and MOS spectra of each
sub-sample separately in order to check the consistency between the PN and MOS
stacked spectra. The results are also included in
Table\,\ref{tab:fit_subsamples}. In each sub-sample, the line EWs estimated
from the PN and MOS spectra are in good agreement with each other while the
photon indices are consistent within 90 per cent confidence level (the typical
90 per cent confidence level of $\Gamma$ is 0.06).

%%*************************************************
%%**************Section: Discussion****************
%%*************************************************
\section{Discussion}\label{sec:discussion}

\subsection{\label{subsec:IT_relation}On the X-ray Baldwin effect in type-1 AGN}

\begin{figure*} 
    \begin{center}
    \includegraphics[width=\textwidth]{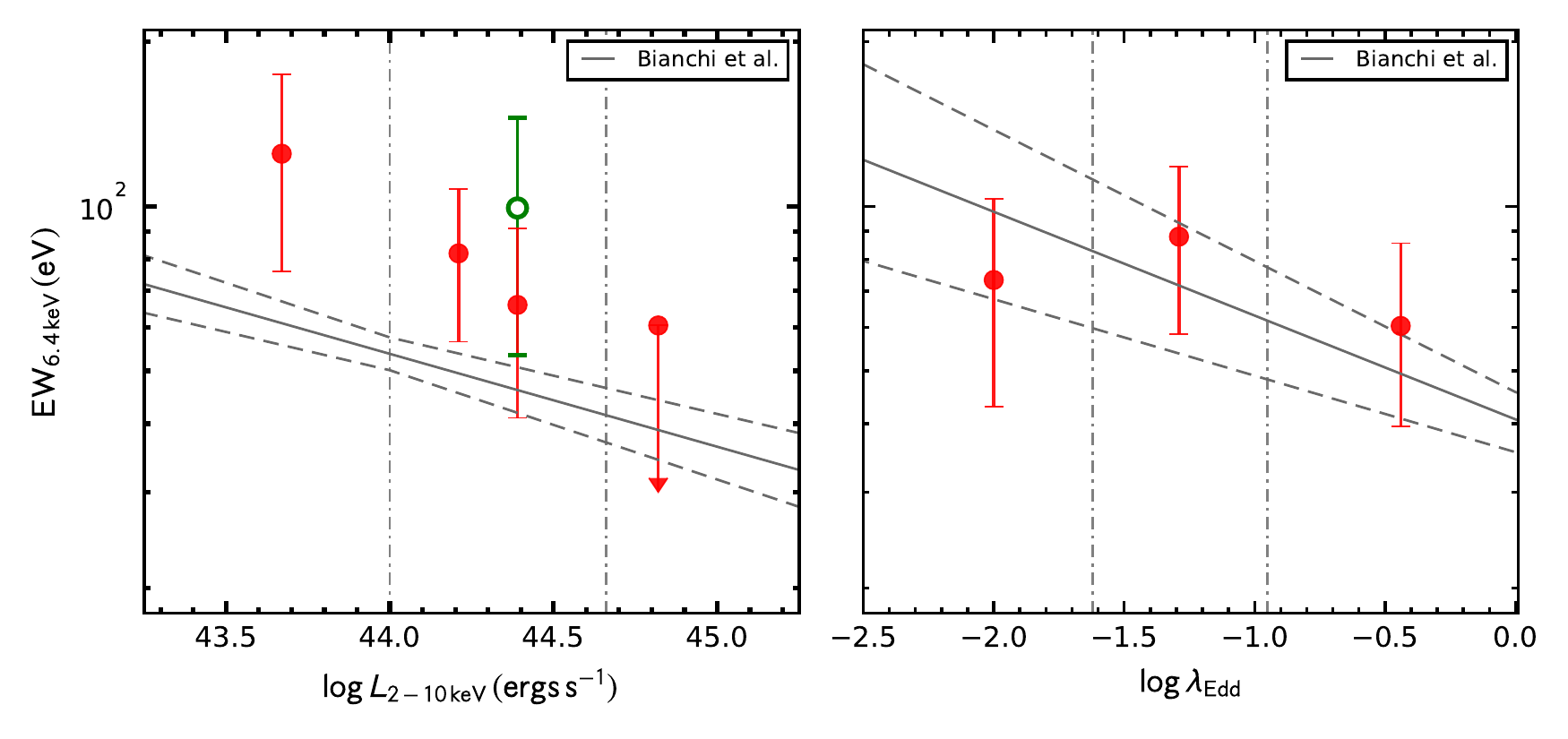}
    \caption{\label{fig:para_vs_ew}Left: the X-ray luminosity versus EWs of Fe
    K$\alpha$ line in the redshift-luminosity sub-samples. The green point
    shows the EWs of the Fe K$\alpha$ line estimated with the base-line only
    (excluding the high ionization line feature) for the HM sub-sample.  Right:
    the $\ledd$ versus EWs of Fe k$\alpha$ line in the $\ledd$ sub-samples. The
    relations presented in \citet{Bianchi+al+2007} as well as the $1\sigma$
    uncertainties are also shown with solid and dashed lines, respectively. The
    vertical dash-dotted lines indicate the luminosity boundaries for the
    sub-samples.}
\end{center} 
\end{figure*}

The left panel of Fig.\,\ref{fig:para_vs_ew} shows the EWs of the Fe K$\alpha$
line against the median $2-10$\,keV X-ray luminosity in the redshift-luminosity
sub-samples. The EWs of the Fe K$\alpha$ line as well as the $1\sigma$
uncertainties are calculated from the spectral modelling as described in
Section\,\ref{sec:xray_prop}. The anti-correlation found in previous work,
$\log\mathrm{EW_{6.4\,keV}}=(1.73\pm0.03)+ (-0.17\pm0.03)\log
L_\mathrm{2-10\,keV,44}$ \citep{Bianchi+al+2007}, as well as its $1\sigma$
uncertainty are also shown as solid and dashed grey lines, respectively.
The two values of the EWs for the HM sub-sample, which are shown as red and
green points, are calculated assuming different models (see
Section\,\ref{subsec:results} above). It shows a trend that the EW of the
narrow Fe K$\alpha$ line decreases with the X-ray luminosity, i.e. the
so-called X-ray Baldwin effect. The anti-correlation was first found by
\citet{Iwasawa+Taniguchi+1993} and confirmed by subsequent studies
\citep[e.g.][]{Nandra+al+1997,Page+al+2004,Jiang+al+2006,Bianchi+al+2007,Shu+al+2010,
Chaudhary+al+2010, Shu+al+2011, Falocco+al+2013, Ricci+al+2014}. However, using a sample of
quasars selected from the Palomar-Green (PG) Bright Quasar Survey,
\citet{Jimenez+al+2005} questioned the significance of the effect. They found
that the EW of the Fe K$\alpha$ line is more likely to be constant for
$\lxray\lesssim10^{45}{\rm erg\,s^{-1}}$ and a step-wise decrease above this
threshold, though they could not rule out a smooth power-law dependence.
\citet{Krumpe+al+2010} argued that the EWs of the Fe K$\alpha$ line in their
sample (three sources with $\lxray>4\times10^{44}{\rm erg\,s^{-1}}$) were
systematically higher than the expected value from the X-ray Baldwin effect
\citep[see also][]{Chaudhary+al+2010}. On the contrary, our results suggest
that the average EWs of the medium (i.e. $44.00<\log\lxray<44.66$) and high
luminosity (i.e. $44.66<\log\lxray<45.60$) bins are still consistent with the
X-ray Baldwin relation found in previous
studies\citep[e.g.][]{Bianchi+al+2007}. We caution that, as remarked by
\citet{Jimenez+al+2005} and \citet{Jiang+al+2006}, these differences may be
potentially due to the contribution from radio-loud sources, which generally
show weak Fe K$\alpha$ emission line \citep{Reeves+Turner+2000}.  However, they
are unlikely to dominate the stacked spectra considering that only
$\sim15$\,per cent of AGN are radio-loud \citep{Urry+Padovani+1995}.

\begin{figure*}
\begin{center}
    \includegraphics[width=\textwidth]{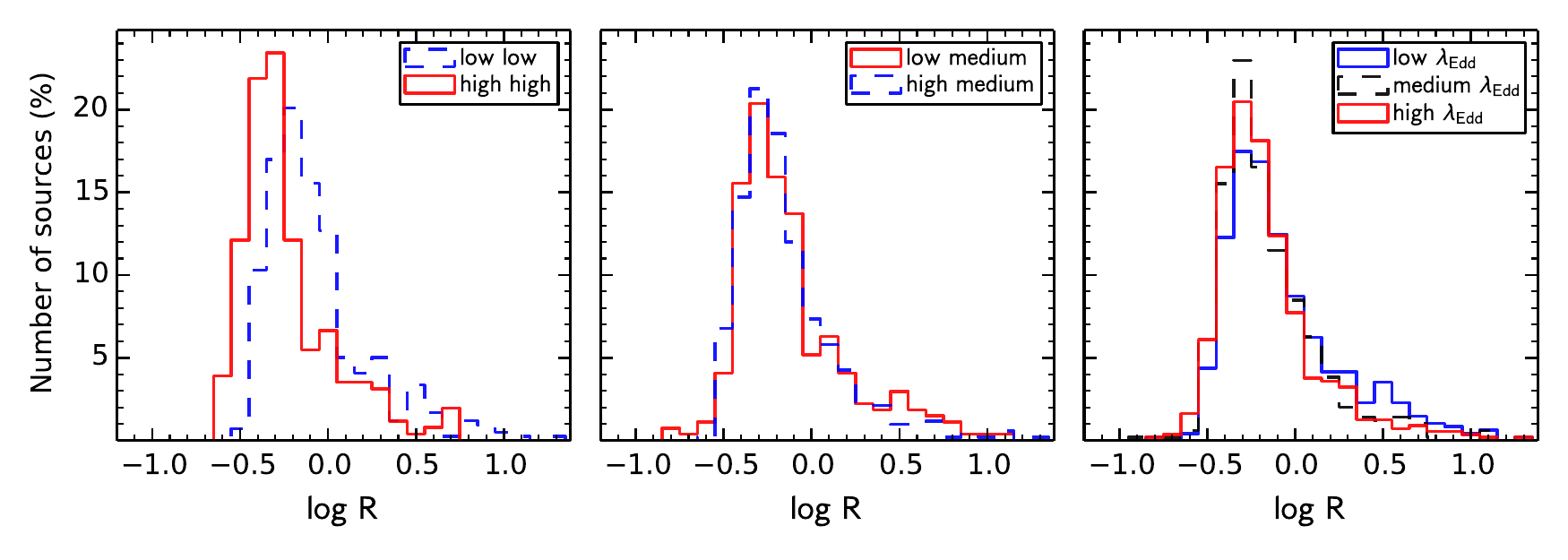}
    \caption{\label{fig:ref_dist} The distributions of the relative reflection
    normalization for the different sub-samples. The reflection normalization, which
    is equal to the reflection fraction, is estimated with the BXA method.}
\end{center}
\end{figure*}

The origin of the X-ray Baldwin effect, which may provide important insights
into the physics of AGN emission, is still unclear. One possible
explanation is the decrease of the covering factor of the torus with X-ray
luminosity \citep[e.g.][]{Page+al+2004, Zhou+Wang+2005}. \citet{Ricci+al+2013a}
showed that the observed X-ray Baldwin effect and its slope can be produced
with this explanation. To provide an independent test of this scenario,
Fig.\,\ref{fig:ref_dist} (left and middle panels) shows the reflection fraction
distribution of the low/high-luminosity and low/high-redshift sub-samples. We
also plot the reflection fraction distribution of the three $\ledd$ sub-samples
in the right panel of Fig.\,\ref{fig:ref_dist}. There is an indication that the
low-luminosity sub-sample has statistically higher reflection fraction than the
high-luminosity sub-sample. A Kolmogorov-Smirnov (K-S) test shows a significant
difference between these two sub-samples ($P_{KS}=1.0\times10^{-17}$),
while the probability that the low/high medium sub-samples are drawn from the same
parent distribution is 60 per cent. The reflection distributions of the medium and
the high $\ledd$ sub-samples are different from the low $\ledd$ sub-sample ($P_{KS}<3\times10^{-4}$),
while no significant difference is found between the these two sub-samples($P_{KS}=0.77$). These
results, which are consistent with the results obtained from the spectral
stacking, may suggest that the X-ray Baldwin effect is more related to the
X-ray luminosity and unlikely to be due to the $\ledd-\Gamma$ relation
\citep[e.g.][ see right panel of Fig. \ref{fig:para_vs_ew}]{Ricci+al+2013b}. Our results also indicate that the structure of
the dusty torus in type-1 AGN does not evolve strongly with redshift. The
difference of the reflection fraction distribution in the low and high
luminosity sub-samples supports the luminosity-dependent covering factor
scenario, i.e. the covering factor of the torus decrease with the X-ray
luminosity.

The X-ray Baldwin effect can also be explained by a luminosity-dependent
ionization state of the emitting material \citep{Nandra+al+1997,Nayakshin+2000}
or by AGN variability \citep{Jiang+al+2006, Shu+al+2012}. In the first case, we
may expect the change of ionization state of the Fe line in different
sub-samples. However, a highly ionized Fe K line (Fe\,{\sc xxvi}) is unambiguously
detected only in the LM sub-sample.  No clear trend of the change of ionization
state with X-ray luminosity is found in our results. Our analysis does not
support the luminosity-dependent ionization state scenario. Regarding
variability as a possible origin, with a large number of AGN in our sample,
which should include sources in different luminosity states in each sub-sample,
we should expect a nearly constant Fe K$\alpha$ EW in different
sub-samples, if the X-ray Baldwin effect can be purely explained by AGN
variability. Our results seem to disfavour this interpretation.

\subsection{\label{subsec:gamma_edd}On the correlation between the photon index and Eddington ratio}

\begin{figure} 
    \begin{center}
    \includegraphics[width=\columnwidth]{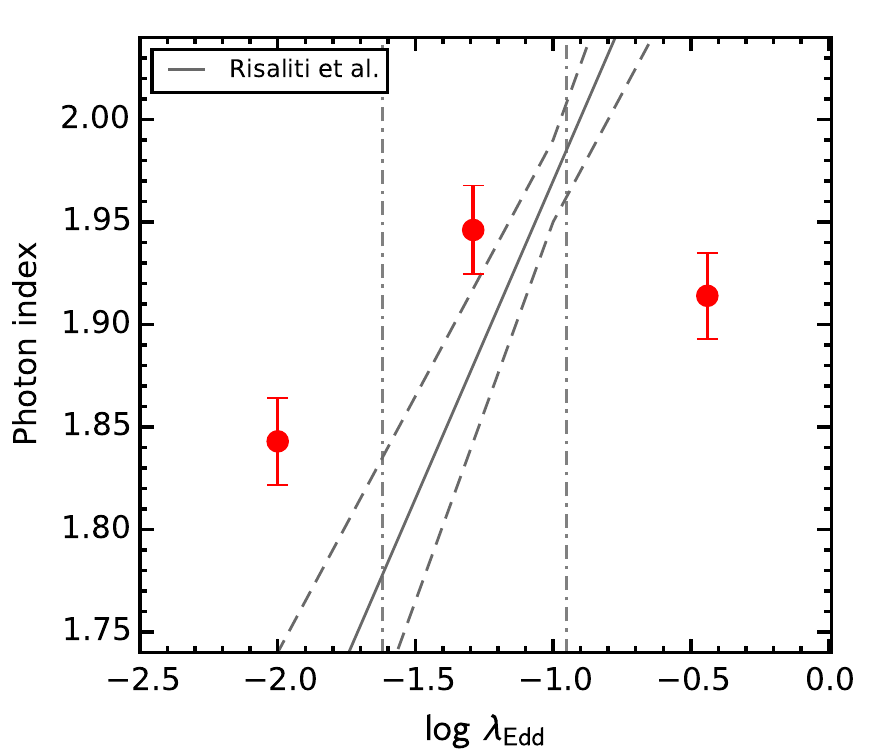}
    \caption{\label{fig:edd_vs_phoind} Photon index vs. $\ledd$ for three
    $\ledd$ subsamples. The solid and dashed lines show the relation as well as
    the 1$\sigma$ uncertainties in \citet{Risaliti+al+2009}, respectively. The
    vertical dash-dotted lines indicate the $\ledd$ boundaries for the three
    sub-samples.}
\end{center} 
\end{figure}

A significant positive correlation between the photon index $\Gamma$ and
$\ledd$ has been found in many previous studies \citep[e.g.][]{Lu+Yu+1999,
Wang+al+2004, Shemmer+al+2006, Shemmer+al+2008, Risaliti+al+2009,
Brightman+al+2013}.  In Fig.\,\ref{fig:edd_vs_phoind} we show the best-fitting
$\Gamma$ against the median values of $\ledd$ for the Eddington ratio
sub-samples. The solid and dashed lines show the relation as well as the
1$\sigma$ uncertainties found in \citet{Risaliti+al+2009}, respectively.
Unlike the strong correlation presented in the studies mentioned above, our
results suggest that the $\Gamma$ stays constant or decreases with $\ledd$ if
$\ledd$ is high, e.g. $\ledd>0.1$. \citet{Ai+al+2011} suggested that the
$\Gamma$ of the sources in their sample, which are narrow-line Seyfert 1
galaxies (NLS1s) with high accretion rate, are systematically below the
\citet{Risaliti+al+2009} relation. This flattening of the $\Gamma-\ledd$ relation was
also confirmed by \citet{Kamizasa+al+2012} with a sample which have small
$\mobh$ and high $\ledd$. Our result, which also shows a flattening of the
relation, is consistent with these studies.

\citet[][see also \citealt{Constantin+al+2009, Younes+al+2011}]{Gu+Cao+2009}
found an anti-correlation between the $\Gamma$ and $\ledd$ in a sample of
low-luminosity AGN (LLAGN), which typically have low $\ledd$ (i.e.
$\ledd<10^{-3}-10^{-2}$). Based on this anti-correlation they proposed that the
X-ray emission of LLAGN could originate from the Comptonization process in
advection-dominated accretion flow (ADAF), in which case the flattening of
$\Gamma$ is due to the increase of the Compton-$y$ parameter with Eddington
ratio.  The positive correlation, generally shown in sources with
$\ledd>10^{-3}-10^{-2}$ \citep{Younes+al+2011}, could be explained by a
disc-corona system.  In this case, the $2-10{\rm keV}$ X-ray emission is
generally thought to be due to the Compton up-scattering of seed soft photons
from  a standard optically thick geometrically thin accretion disc
\citep{Shakura+Sunyaev+1973} in a hot corona
\citep{Haardt+Maraschi+1991, Haardt+Maraschi+1993}. The steepening of $\Gamma$
can be interpreted as enhanced emission from the accretion disc more
effectively cooling the corona as $\ledd$ increases \citep{Pounds+al+1995,
Haardt+Maraschi+1991, Haardt+Maraschi+1993, Wang+al+2004}. If the standard
accretion disc model and the disc-corona system is still appropriate in the
high $\ledd$ regime, the flattening of the $\Gamma-\ledd$ relation, as
presented in our result \citep[see also][]{Ai+al+2011, Kamizasa+al+2012},
suggest that the inverse Compton scattering is saturated in sources with high
$\ledd$.  On the other hand, the structure of the accretion disc may change if
the $\ledd$ is high, i.e. a slim disc \citep{Abramowicz+al+1988} instead of the
standard thin disc. If this is the case, the flattening of the relation can
also be explained under the assumption that the temperature of the slim disc
does not change anymore when $\ledd$ exceeds a certain value. Thus increasing
the $\ledd$ will not produce more soft X-ray photons, leading to a flatter
$\Gamma-\ledd$ relation. We caution here that the uncertainties in the $\mobh$
measurements, which may suffer from many systematic effects such as orientation
\citep{Shen+2013, Shen+Ho+2014}, can potentially destroy the $\Gamma-\ledd$
correlation, especially in the high $\ledd$ sub-sample which consist of a large
fraction of sources using the less reliable C\,IV and Mg\,II based virial BH
masses.

As mentioned in \citet{Iwasawa+al+2012}, the flattening of the $\Gamma$ in the
high $\ledd$ stacked spectrum may potentially be due to absorption. This is
unlikely the case in our sample as no significant difference of the column
density distribution of the three $\ledd$ sub-samples is found.

\subsection{\label{subsec:non_bl}Non-detection of the broad line in the BH mass
sub-samples?}

A broad line profile was seen in the previous X-ray spectral stacking studies
of AGN \citep[e.g.][]{Streblyanska+al+2005, Corral+al+2008, Iwasawa+al+2012,
Falocco+al+2014, Liu+al+2015}, although typically with low significance.  The
broad Fe K$\alpha$ line, which is generally thought to be broadened due to the
strong gravitational field of the central SMBH, can be used to constrain the
spin of BH \citep{Brenneman+Reynolds+2006, Dauser+al+2010}.  The SMBH growth is
mainly due to accretion of matter during active phases over cosmological
times. For a standard optical thick geometry thin accretion disc, the radiation
efficiency depends solely on the spin of BH. Thus we may expect a dependence of
the $\mobh$ growth rate on the spin of the BH. \citet{Volonteri+al+2013} showed
that spin as a function of $\mobh$ in galaxies under different accretion
scenarios (i.e. coherent or chaotic accretion) are different. In principle, by
comparing the line profile detected in the stacked spectra of different $\mobh$
samples may help us understand the accretion history of the BH.  However, as
shown in Fig.\,\ref{fig:edd_mbh_subs}, no apparent broad line features are
detected in the three $\mobh$ sub-samples. This should not be a surprise since
the S/N of the stacked spectra is still too low to detect the broad line
\citep[e.g. total photon counts $>1.5\times10^5$ is required to detect the
broad line,][]{deLaCalle+al+2010}. A large sample of AGN with well measured
$\mobh$ from optical spectra (e.g. SDSS IV/SPIDERS) will be helpful to
understand the relation between BH spin and $\mobh$.

\subsection{\label{sec:x_opt_class}On the inconsistency between X-ray/optical
classifications}

As shown in Fig.\,\ref{fig:nh_dist}, the \nh distribution of type-1 AGN is
peaked at values (e.g. $\sim20.7$) consistent with little or no X-ray absorption
while most of the type-2 AGN have \nh larger than $10^{21}~{\rm cm^{-2}}$.
However, about $\sim40$\,per cent of both type-1 and type-2 AGN show
discrepancies between their optical classifications and the absorption column
densities obtained from the X-ray spectral fitting, i.e. optical type-1 AGN with
indications of high X-ray absorption or type-2 AGN with X-ray spectra showing
no significant absorption. 

\begin{figure}  
    \begin{center}
    \includegraphics[width=\columnwidth]{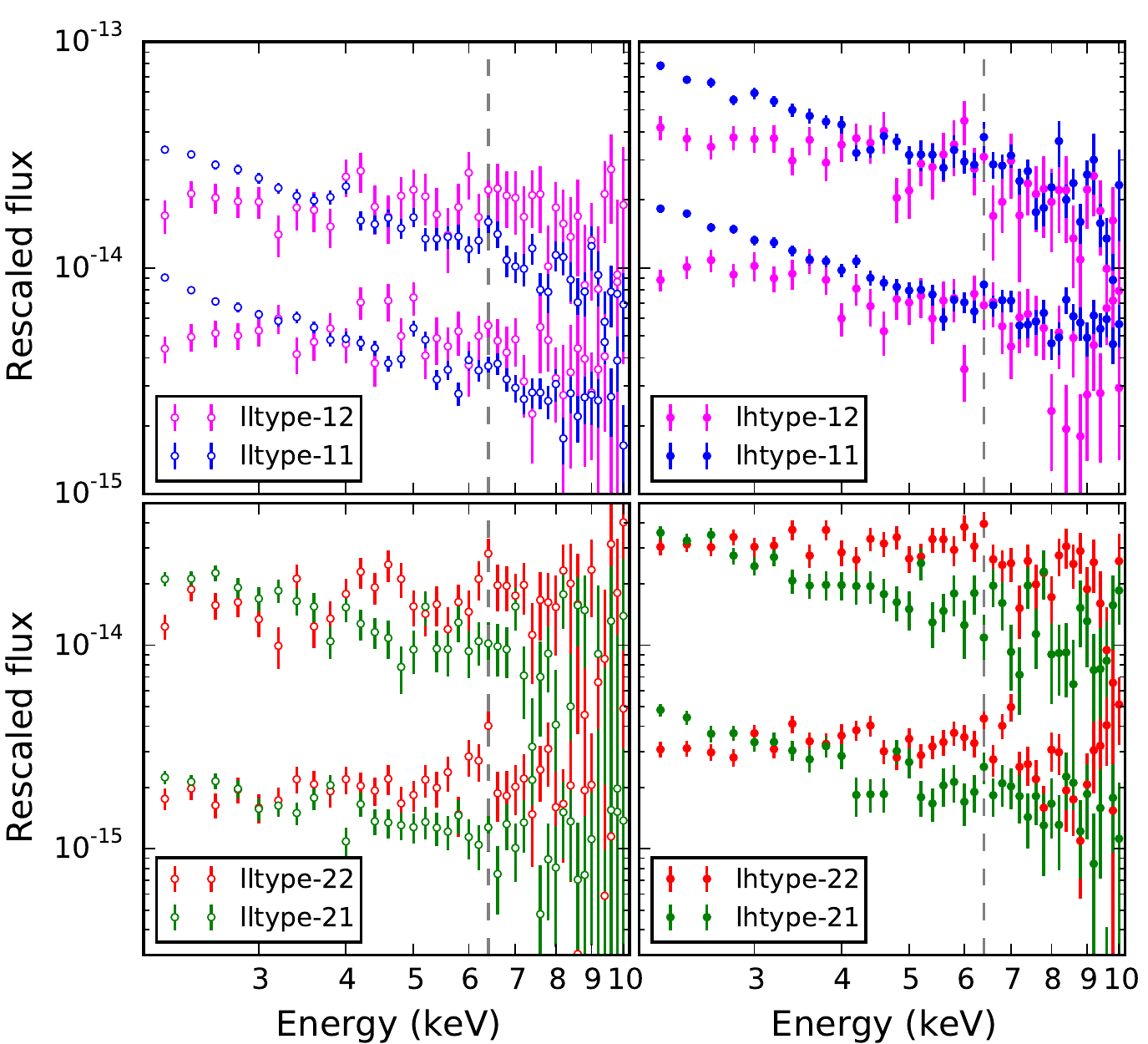}
    \caption{\label{fig:low_red_nh} {\it Upper-left} panel: the stacked spectra
	of the PN (with higher rescaled flux) and MOS (with lower rescaled flux)
	spectra for the low-luminosity low-redshift optical classified type-1 AGN
	with (open magenta, {\it lltype-12}) and without (open blue, {\it
	lltype-11}) X-ray absorption signature. {\it Upper-right} panel: the same
	as the upper-left panel but with high-luminosity ones (filled magenta, {\it
	lhtype-12}; filled blue, {\it lhtype-11}). {\it Bottom-left} panel: same as
	above but for the low-luminosity low-redshift optical classified type-2 AGN
	with (open red, {\it lltype-22}) and without (open green, {\it lltype-21})
	X-ray absorption signature. {\it Bottom-right} panel: the same as the
	bottom-left panel but with high-luminosity ones (filled red, {\it lhtype-22};
	filled green, {\it lhtype-21}). The grey dashed vertical line indicates the
	6.4~keV emission line.}
\end{center} 
\end{figure}

\begin{figure}  
    \begin{center}
    \includegraphics[width=\columnwidth]{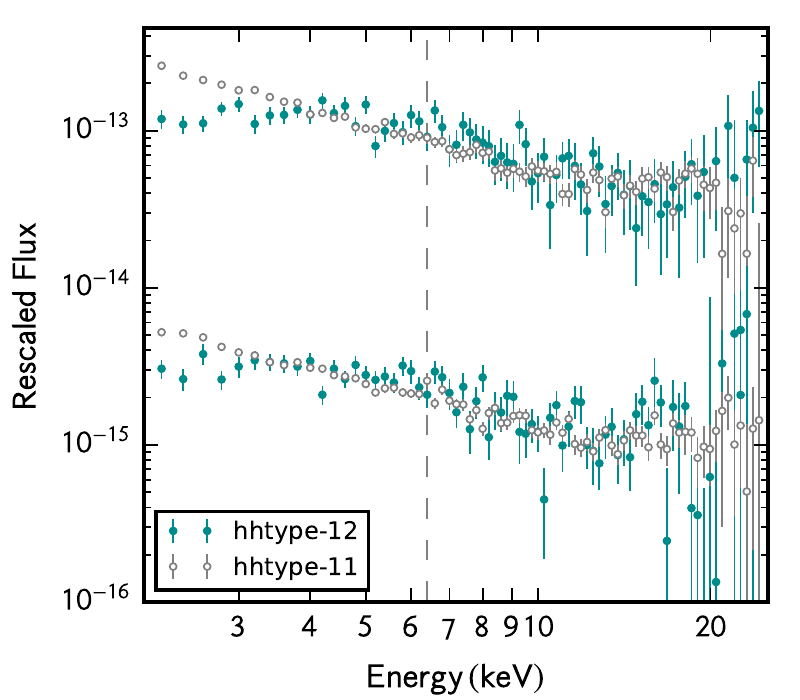}
	\caption{\label{fig:hig_red_nh}The 2-25\,keV PN (with higher rescaled flux) and MOS
	(with lower rescaled flux) stacked spectra for the high-redshift high-luminosity
type-11 (hhtype-11, grey points) and type-12 (hhtype-12, cyan points).  The
grey dashed vertical line indicates the 6.4~keV emission line.}
\end{center} 
\end{figure}

\begin{figure}  
    \begin{center}
    \includegraphics[width=\columnwidth]{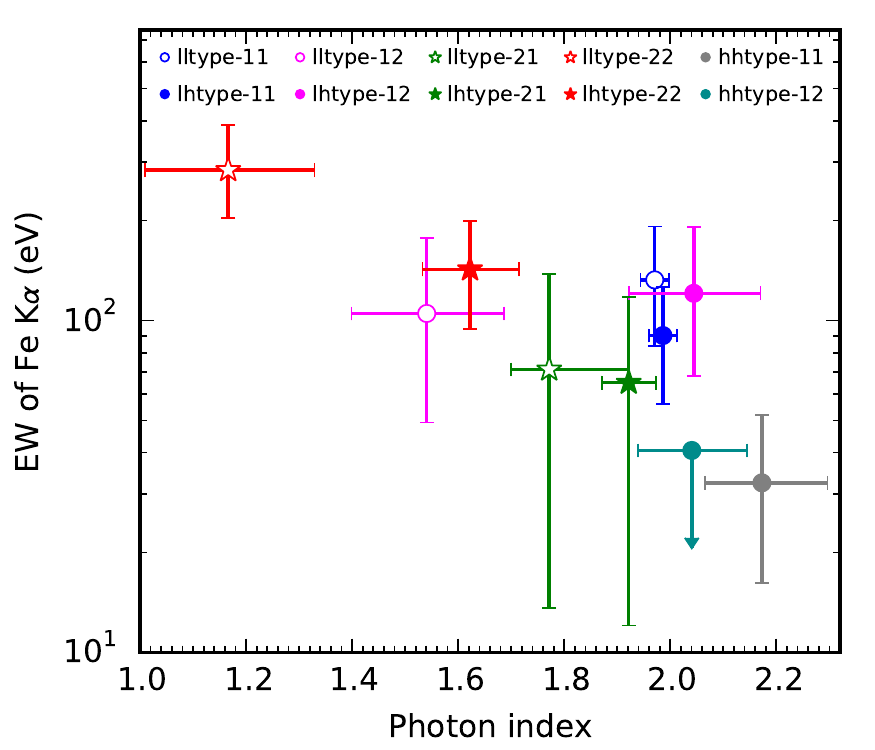}
    \caption{\label{fig:para_nh} The equivalent width of the Fe K$\alpha$ line
    versus the photon index. The optical type-1 AGN are shown with empty/filled circles while
    type-2 are shown with empty/filled stars.}
\end{center} 
\end{figure}

Following \citet{Merloni+al+2014}, sources with $N_{\rm H}>10^{21.5}~{\rm
cm^{-2}}$ are considered to be X-ray obscured. We define type-11 AGN as those
which are optically classified as type-1 AGN with unobscured X-ray spectra;
type-22 AGN are instead optical type-2 objects which are also obscured at
X-ray. Type-12 AGN are type-1 objects which are X-ray obscured; conversely we
call type-21 as optical type-2 AGN which have X-ray spectra consistent with no
obscuration. Each of these types are divided into different bins on the basis
of the redshift and X-ray luminosity. The high-redshift high-luminosity
type-11/12 (hhtype-11/12) bin include all the type-11/12 sources with $z>1.36$
and $\log\lxray>44~{\rm erg~s^{-1}}$. The low-redshift ($z<1.36$) type-11/12
sources are divided into two luminosity bins, i.e. low-luminosity bin with
$\log\lxray$ in the range $42-44~{\rm erg~s^{-1}}$ (lltype-11/12) and high
luminosity bin with $\log\lxray$ larger than $44~{\rm erg~s^{-1}}$
(lhtype-11/12). The same procedure is applied to the low-redshift ($z<1.36$)
type-21/22 sources but with different luminosity range, i.e. lltype-21/22 with
$\log\lxray$ in the range $42-43~{\rm erg~s^{-1}}$ and lhtype-21/22 with
$\log\lxray$ in the range $43-44~{\rm erg~s^{-1}}$.

To investigate the properties of these different types, the same X-ray spectral
stacking method described in Section\,\ref{sec:spec_stack} is applied to each bin.
In Fig.\,\ref{fig:low_red_nh}, we show the stacked spectra of all the low
redshift bins. The stacked spectra of the hhtype-11/12 are plotted in
Fig\,\ref{fig:hig_red_nh}. We fit the stacked spectra with a model including a
photoionization absorption component, a power-law continuum, and, if necessary,
a Gaussian component ({\sc phabs*[po+gaussian]}). Some of the best-fitting
parameters as well as the EWs of the Gaussian component are listed in
Table\,\ref{tab:nh_fit}. The measured column densities of the stacked spectra
are well in agreement with the values of individual source in each bin, i.e.
bins with \nh higher/less than $10^{21.5}~{\rm cm^{-2}}$ show X-ray absorption
larger/smaller than $\sim10^{22}~{\rm cm^{-2}}$. This implies that the column
density of an individual source obtained through the spectral modelling method,
though with large uncertainties due to the low S/N, are statistically reliable. 

\begin{table}
\bgroup
\setlength\tabcolsep{6.0pt}
\begin{center}
\caption{\label{tab:nh_fit}Estimated parameters for the \nh samples}
\begin{tabular}{lcccc}\\\hline\hline
   samples&   $\Gamma$    &\nh $(10^{22}~{\rm cm^{-2}})$& $EW_{6.4~{\rm keV}}$ (eV) & \#src \\\hline
lltype-11 & $1.97\pm0.03$ &            $<0.08$          &  $133^{+59}_{-49}$ & 363 \\
lhtype-11 & $1.99\pm0.03$ &            $<0.06$          &  $ 90\pm36 $       & 251 \\
hhtype-11 & $2.17\pm0.13$ &         $0.20\pm0.16$       &  $ 32^{+20}_{-16}$ & 662 \\
lltype-12 & $1.54\pm0.14$ &         $1.56\pm0.60$       &  $105^{+72}_{-56}$ & 55 \\
lhtype-12 & $2.04\pm0.13$ &         $1.66\pm0.49$       &  $121^{+70}_{-53}$ & 38 \\
hhtype-12 & $2.04\pm0.10$ &         $2.97\pm0.39$       &  $<40.6$           & 111 \\
lltype-22 & $1.17\pm0.16$ &         $0.93\pm0.59$       & $284^{+105}_{-81}$ & 66\\
lhtype-22 & $1.62\pm0.09$ &         $1.83\pm0.38$       &  $143^{+57}_{-49}$ & 118 \\
lltype-21 & $1.77^{+0.15}_{-0.07}$ &   $<0.52$          &  $ 71^{+67}_{-57}$ & 127 \\
lhtype-21 & $1.92\pm0.05$ &            $<0.30$          &  $ 65^{+73}_{-53}$ & 134 \\\hline
\end{tabular}
\parbox[]{8.5cm}{lltype-11/lhtype-11: low-redshift low/high-luminosity X-ray unobscured optical type-1 AGN;
				  hhtype-11: high-redshift high-luminosity X-ray unobscured optical type-1 AGN;
				  lltype-12/lhtype-12: low-redshift low/high-luminosity X-ray obscured optical type-1 AGN;
				  hhtype-12: high-redshift high-luminosity X-ray obscured optical type-1 AGN;
				  lltype-21/lhtype-21: low-redshift low/high-luminosity X-ray unobscured optical type-2 AGN;
				  lltype-22/lhtype-22: low-redshift low/high-luminosity X-ray obscured optical type-2 AGN;
				}
\end{center}
\egroup
\end{table}

\subsubsection{X-ray obscured type-1 AGN}

The {\it upper-left} panel of Fig.\,\ref{fig:low_red_nh} shows the stacked
spectra of both the X-ray unobscured ({\it open blue} points) and obscured
({\it open magenta} points) low-redshift low-luminosity optical type-1 AGN
(lltype-11/12).  The {\it upper-right} panel of Fig.\,\ref{fig:low_red_nh}
shows the same plot but for the high-luminosity ones (lhtype-11/12). As clearly
demonstrated in the plots, both the lltype-11 and lhtype-11 AGN exhibit a featureless
power-law with similar photon index (see upper panel of
Fig.\,\ref{fig:low_red_nh}, with open blue for lltype-11 and filled blue for
lhtype-11). The EW of the Fe line seems to be weaker (though with large errors,
see Table\,\ref{tab:nh_fit}) in the lhtype-11 AGN,
which is consistent with the X-ray Baldwin effect and the results presented in
Section\,\ref{subsec:IT_relation}. The stacked spectra of the lltype-12 and
lhtype-12 AGN are quite different, except that both show significant
obscuration in their stacked X-ray spectra. The photon index of the lhtype-12 bin is
$\Gamma=2.04\pm0.13$ which is consistent with the typical value found in type-1
AGN. In contrast, the lltype-12 AGN has a significantly shallower slope, i.e.
photon index $\Gamma=1.54\pm0.14$. As shown in Fig.\,\ref{fig:hig_red_nh}, the
hhtype-11 has a overall similar stacked spectrum to the lltype-11 as well as
the lhtype-11, except for a relatively weak Fe K$\alpha$ line. Even the X-ray
obscured ones, i.e. the hhtype-12 and lhtype-12, does {\it not} show
significant differences. They both have a photon index $\Gamma\sim2.04$ and
a weak Fe K$\alpha$ emission line, although a highly ionized emission line is
detected in the stacked spectra of hhtype-12.

The estimated $N_{\rm H}$ from low S/N X-ray spectra of unobscured type-1 AGN
can be higher than $10^{21}~{\rm cm^{-2}}$, as shown in the simulations by
\citet{Krumpe+al+2008}. However, only $\sim2$\,per cent of the sources in the
simulations (the sources in those simulations have 40 net photon counts with
$z=1$) showed $N_{\rm H}$ higher than $10^{21.5}~{\rm cm^{-2}}$, which is much
smaller than the fraction (i.e. $>20$\,per cent, $\sim10$\,per cent if only
sources with net photon counts larger than 50 are selected) of the X-ray obscured
sources in our sample. We thus conclude that the majority of the type-12 AGN in
our sample are truly X-ray obscured. \citet{Merloni+al+2014} suggested that
the excess X-ray absorption presented in their high X-ray luminosity high
redshift type-12 AGN ($\lxray>10^{44}~{\rm erg~s^{-1}}$, $1.5<z<2.5$) may be
produced by dust-free gas within (or inside) the broad-line region. This
suggestion was strengthened in a recent work by \citet{Davies+al+2015}. Using a
complete volume limited hard X-ray selected AGN sample, the authors argued that
at high-luminosity (e.g. $\log{L_{14-195}}>44$) the torus consist of two parts,
an outer part which is the dusty torus and an inner dust-free part which is
responsible for the excess X-ray obscuration. Our work seems to be consistent
with this scenario. As shown in Fig.\,\ref{fig:hig_red_nh}, the rest-frame $4-25\,{\rm keV}$ 
stacked spectra of the hhtype-11 and 12 AGN are almost identical
(this seems to also be true for the lhtype-11/12 AGN, though only the
spectra below $\sim10\,{\rm keV}$ are available). On average, the
hhtype-12 AGN do {\it not} show a stronger reflection component than the
hhtype-11, i.e. $R_{\rm refl}<1.7$ for hhtype-12 and $R_{\rm refl}=2.25^{+0.72}_{-0.58}$ for
hhtype-11 if we fit the stacked spectra with the {\tt pexrav} model.
Both the stacked spectra of the high-luminosity type-12 sub-samples have
steep slopes ($\Gamma\sim2.04$), which is consistent with the slopes
found in typical type-1 AGN, showing no evidence for strong reprocessing of the
primary continuum.
Moreover, the EWs of the Fe K$\alpha$ line in the high-luminosity type-12 sub-samples
are weaker than that found in typical type-2 AGN,
e.g. $\gtrsim200$\,eV \citep{LaMassa+al+2009}. These results may imply that the
high-luminosity type-12 AGN have similar torus structure to the typical type-1 AGN.
Thus the excess X-ray absorption in the high-luminosity type-12 AGN is more likely
to be due to dust-free gas which may probably locate close to or within the BLR,
instead of the dusty torus. In addition, a $\sim6.67$~keV (i.e $E=6.67\pm0.05$
if fitted with a Gaussian model) emission line, which must be the Fe\,{\sc xxv}
line, is detected in the stacked spectra of hhtype-12 AGN. This suggest that,
when the luminosity is high, the dust-free gas can be ionized by the strong
radiation from the central region.

On the contrary, the shallow slope of the power-law continuum as well as
the EW of the Fe K$\alpha$ line (see Fig.\,\ref{fig:para_nh}, open magenta
circle) are indications that the average X-ray properties of the sources in the
low-redshift low-luminosity type-12 (lltype-12) bin are close to the typical
type-2 AGN. These results thus suggest that the excess X-ray absorption in the
lltype-12 AGN is likely to be due to obscuration by the torus. One possibility
is that the low-luminosity sources have on average higher torus covering factor
than the high-luminosity ones, i.e. luminosity-dependent dust covering factor
as suggested in Section\,\ref{subsec:IT_relation}. This is also in accord with
\citet{Davies+al+2015}, in which have shown that the opening angle of the dusty
torus is smaller at lower luminosity (e.g. $\log{L_{14-195}}<44$). The
detection of the broad emission line in the optical spectra either suggest that a clumpy
torus structure is common in AGN \citep[e.g.][]{Coffey+al+2014}, or could be
due to variability obscuration and the non-simultaneous of X-ray/optical
observations.

\subsubsection{X-ray unobscured type-2 AGN}

The stacked spectra for the low/high-luminosity type-2 AGN are shown in the
bottom panel of Fig.\,\ref{fig:low_red_nh}. Both the lltype-22/lhtype-22 AGN,
which are marked as open/filled red stars in Fig.\,\ref{fig:para_nh}, have
X-ray properties of the typical type-2 AGN, i.e.  shallow power-law photon
indices ($1.17\pm0.16$ for lltype-22 and $1.62\pm0.09$ for the lhtype-22, the
abnormal low photon index of the lltype-22 can be potentially due to the
degeneration with column density) and relatively high EWs of Fe K$\alpha$ line
($284^{+105}_{-81}$ for lltype-22 and $143^{+57}_{-49}$ for lhtype-22). The
difference of the EWs of the Fe K$\alpha$ line between the two luminosity bins
may imply that the fraction of heavily obscured sources in the low-luminosity
type-22 is higher than the high-luminosity ones, i.e. the obscuration fraction
decrease with X-ray luminosity \citep[e.g.][and references therein]{Merloni+al+2014}.

Only upper limits of the column density ($\sim5\times10^{21}~{\rm cm^{-2}}$)
can be given for the lltype-21 and lhtype-21 AGN.  They have very similar X-ray
properties, both the EW of Fe K$\alpha$ line as well as the photon index are
consistent well within uncertainties.

There are several reasons which may possibly explain the lack of X-ray
obscuration in these optical type-2 AGN. The first one is that, due to the low
S/N of the X-ray spectra, the \nh of most sources in the type-21 may actually
be underestimated. However, the relatively weak Fe K$\alpha$ line seems to be
inconsistent with this explanation. Furthermore, we still have about 40 per
cent of X-ray unobscured type-2 AGN even when only sources with higher S/N are
selected. It is also possible that these X-ray unobscured type-2 AGN are
contaminated by type-1 sources. In \citet{Davies+al+2015}, they concluded that
the X-ray unabsorbed type-2 AGN are rare and the high fraction of type-21 found
in \citet{Merloni+al+2014} is due to a bias in the photometric classification.
In our sample, the type-2 sample we used include sources which are classified
as type-2 candidate (NLAGN2cand) in \citet{Menzel+al+prep}. These type-2
candidates mainly consist of low (optical spectroscopic) S/N type-2 AGN, but
may also include contaminating star-forming galaxies as well as type-1 AGN. The
fraction of type-21 AGN in these sources ($\sim55$ per cent) is higher than
that in the pure type-2 sample ($\sim40$ per cent). In addition, both the pure
type-2 and the type-2 candidate sample may include the so-called true type-2
AGN \citep[e.g.][see also \citealt{Shu+al+2007, Valencia+al+2012, Ichikawa+2015} and references therein]{Panessa+Bassani+2002, Brightman+Nandra+2008, Shi+al+2010,
Bianchi+al+2012, Li+al+2015}, i.e. AGN without any scattered or hidden BLR signature. The optical type-2 AGN without X-ray obscuration feature can also be
explained by including the so-called elusive AGN (eAGN-SFG and eAGN-ALG, see
Section\,\ref{subsec:opt_info}). They can be weak type-1 AGN for which the
optical spectra are dominated by the host galaxies.  The X-ray column density
distribution, with about 75 per cent of these elusive AGN have \nh less than
$10^{21.5}~{\rm cm^{-2}}$, seems to support the suggestion that these are more
likely to be weak type-1 AGN.  Due to relatively small number of type-2 AGN in
our sample, we cannot further split the type-21 into different sub-samples
(e.g. pure type-2, type-2 candidate and elusive AGN) to investigate which is
the main contributor of the X-ray unobscuration. Further studies of large
samples with high S/N is necessary to resolve the problem.

%%*************************************************
%%****************Section: Summary*****************
%%*************************************************
\section{Summary and conclusion}\label{sec:summary}

In this paper, we studied the X-ray properties of a large AGN sample selected
in the \xxln field. In total, $8445$ point-like X-ray sources are detected in
this field, of which $\sim2512$ have reliable redshift measurements as well as
optical classifications from a dedicated SDSS-III/BOSS ancillary program. Those
sources with redshift measurements are selected as the main AGN sample in this
work. The details of the 8445 X-ray detected sources can be found online, while
the properties of the sources with redshift measurements are presented in
Table\,\ref{tab:xray_cata}. By performing the spectral fitting to the BOSS
spectroscopy of the type-1 objects in our sample, we estimate the single-epoch
virial BH masses from continuum luminosities and broad line widths. To estimate
the bolometric luminosity, we calculated the radiation directly produced by the
accretion process, i.e.  the optical/UV thermal radiation which is computed
assuming a standard thin disc and the hard X-ray radiation estimated from the
X-ray spectral analysis. The $\mobh$, bolometric luminosity, $\ledd$ and other
optical properties for the type-1 objects are described in
Table\,\ref{tab:opt_cata}.

We analysed the X-ray data of our AGN sample by means of X-ray spectral
modelling and X-ray spectral stacking. We adopted the BXA software to fit the
X-ray spectra of each individual source, in which case the posterior
distribution of each parameter can be obtained. In order to investigate how the
X-ray properties of AGN changes with the physical parameters, e.g. the X-ray
luminosity, redshift, $\mobh$, $\ledd$ and column density, a X-ray spectral
stacking method is applied to different sub-samples, which are selected on the
basis of the AGN physical parameters.

The main results of our study are summarized below.
\begin{enumerate}
\item We confirm the well-known X-ray Baldwin effect in our
	luminosity-redshift sub-samples, i.e. a decrease of the equivalent
	width of the narrow Fe K$\alpha$ line with increasing X-ray luminosity.
	Furthermore, we find that the low
	luminosity, low-redshift sub-sample has a systematically higher reflection
	fraction than the high-redshiftlow-redshift, high-luminosity one. These results are in
	good agreement with the view that the X-ray Baldwin effect is due to the
	decrease in the covering factor of the torus with X-ray luminosity.
\item By comparing the photon index measured from the three $\ledd$
	sub-samples, we find evidence for a decreasing or constant photon
	index at high $\ledd$, which is contrary to previous works. We suggest
	that the structure of the accretion disc may change as the Eddington
	ratio approaches unity.
\item Following \citet{Merloni+al+2014},
    We investigated the properties of the type-11/12 AGN sample
	by the X-ray spectral stacking method. We did not find any evidence for a
	higher reflection fraction in the high-luminosity type-12 sample comparing
	with high-luminosity type-11 sample. We thus suggested that the excess
	X-ray absorption shown in those optical broad line AGN can be due to small scale dust-free
	gas within (or close) to the BLR. However, the stacked X-ray spectrum of
	the low-luminosity type-12 sample has a shallower slope than the type-11
	counterpart. This is an indication that the X-ray obscuration in those
	objects may be due to a clumpy torus with large covering factor.
\item The situation of the type-21/22 AGN is more complicated and less clear,
	due to a relatively small number of type-2 AGN in our sample. One possible explanation for
	the lack of X-ray obscuration in the type-21 AGN is the contamination from
	star-forming galaxies, type-1 AGN, true type-2 AGN or the elusive type-2 AGN. Future X-ray
	studies with large spectroscopic type-2 AGN samples, such as those selected
	with eROSITA \citep{Merloni+al+2012}, can help to resolve this problem.
\end{enumerate}

\section{Acknowledgements}

ZL thanks Murray Brightman, Tom Dwelly, Kazushi Iwasawa, Yuan Liu, Youjun Lu,
Xinwen Shu and Weimin Yuan for useful
discussions. Thanks are given to Damien Coffey for a careful read of the manuscript.
ZL is grateful for the support by the MPG--CAS Joint Doctoral
Promotion Programme and the hospitality of MPE, where the research was carried
out. This work benefited from the thales project 383549 that is jointly funded
by the European Union and the Greek Government in the framework of the
programme `Education and lifelong learning'.  Funding for SDSS-III has been
provided by the Alfred P. Sloan Foundation, the Participating Institutions, the
National Science Foundation, and the U.S.  Department of Energy Office of
Science. The SDSS-III web site is http://www.sdss3.org/.

SDSS-III is managed by the Astrophysical Research Consortium for the
Participating Institutions of the SDSS-III Collaboration including the
University of Arizona, the Brazilian Participation Group, Brookhaven National
Laboratory, Carnegie Mellon University, University of Florida, the French
Participation Group, the German Participation Group, Harvard University, the
Instituto de Astrofisica de Canarias, the Michigan State/Notre Dame/JINA
Participation Group, Johns Hopkins University, Lawrence Berkeley National
Laboratory, Max Planck Institute for Astrophysics, Max Planck Institute for
Extraterrestrial Physics, New Mexico State University, New York University,
Ohio State University, Pennsylvania State University, University of Portsmouth,
Princeton University, the Spanish Participation Group, University of Tokyo,
University of Utah, Vanderbilt University, University of Virginia, University
of Washington, and Yale University.

\bibliographystyle{mnras}

%\bibliography{xmm_xxl}

\appendix
\section[]{XMM Observation log}\label{tab:obs_log}
The individual XMM observations used in this paper is listed in
Table\,\ref{tab:obs_log}. In addition to the \xxln/XMM-LSSS 10ks data, deeper
XMM observations within the \xxln footprint are also included.
\begin{table*}
\caption{XMM-XXL log of observations}\label{tab_xmmlog}
\begin{minipage}{150mm}
\begin{center} 
%\scriptsize
	\begin{tabular}{@{}l c c  cc cc cc@{}}
\hline
XMM obsid & RA & DEC & \multicolumn{2}{c}{EPIC-PN}  & \multicolumn{2}{c}{EPIC-MOS1}   & \multicolumn{2}{c}{EPIC-MOS2} \\
          & \multicolumn{2}{c}{J2000} &     filter & exposure (s) &  filter & exposure (s) &  filter & exposure (s) \\\hline
0037980101  &  02:22:45.61 & -03:50:57.3 &  Thin1  &       10178 &  Thin1  &       14560 &  Thin1  &       14552  \\
0037980201  &  02:24:05.66 & -03:50:58.4 &  Thin1  &        8999 &  Thin1  &       13375 &  Thin1  &       13235  \\
0037980301  &  02:25:25.73 & -03:50:57.8 &  Thin1  &        9100 &  Thin1  &       13476 &  Thin1  &       13436  \\
0037980401  &  02:26:45.47 & -03:50:59.1 &  Thin1  &        3680 &  Thin1  &        6020 &  Thin1  &        5340  \\
0037980501  &  02:28:05.38 & -03:51:00.7 &  Thin1  &       12268 &  Thin1  &       16066 &  Thin1  &       16051  \\
\hline
 \end{tabular} 
\begin{list}{}{}
\item Log of  the XMM observations used to  construct the X-ray source
catalogue  in  the  equatorial  XMM-XXL  field.  Listed  are  (1)  XMM
observation  identification  number.   A  ``U''  at  the  end  of  the
observation ID marks ``unscheduled''  XMM observations.  Data taken in
the  XMM mosaic  mode  are  marked by  a  trailing incremental  number
separated   by   the   observations   identification  number   by   an
underscore. (2) Right Ascension of  the XMM obeservation in J2000, (3)
Declication of the XMM obeservation  in J2000, (4) EPIC PN filter used
during the observations, (5) EPIC PN exposure time in secons, (6) EPIC
MOS1 filter used during the  observations, (7) EPIC MOS2 exposure time
in secons, (8) EPIC MOS2 filter used during the observations, (9) EPIC
MOS2 exposure time in secons. The full list can be found online.
\end{list}
\end{center}
\end{minipage}
\end{table*}

\section[]{Table columns}\label{tab:tab_col}
The information of the full catalogue (see Section\,\ref{subsec:opt_obs}) is list here. The full list can be found online.
\subsection{source detection parameters}

uxid (30a): unique x-ray source identification string. the unique name
of the  each source  is a  string which is  composed from  the project
name, followed  by the number of  the pointing on which  the source is
detected or zero in the case of projects with a single xmm observation
and ending with a sequential number.

\noindent 
box\_id\_src  (j): identification  number  of a  source on  individual
pointings or  observations in the case  of projects with  a single xmm
observation.

\noindent 
ra  (d):  right  ascension  in  degrees of  the  x-ray  source  before
correcting for systematic offsets.

\noindent 
dec (d): declination in degrees  of the x-ray source before correcting
for systematic offsets.

\noindent 
radec\_err  (e):  positional  uncertainty  in  arcsec  of  the  source
position. this parameter  is estimated by the {\sc  ewavelet} task of sas. 

\subsection{Source parameters related to flux estimation}

\noindent 
CNT\_TOTAL\_FULL\_80 (J): full  band counts at the source position
from all EPIC cameras estimated within the 80 per cent EEF ellipse.

\noindent 
BKG\_TOTAL\_FULL\_80  (E):  expected number  of  full band  background
counts at the source position from  all EPIC cameras within the 80 per
cent EEF ellipse.

\noindent 
EXP\_TOTAL\_FULL\_80 (E):  full band exposure  time in seconds  at the
source position from  all EPIC cameras. This parameter  is the average
exposure time within the 80 per cent EEF ellipse.

\noindent 
CNT\_PN\_FULL\_80 (J): full band counts  at the source position on the
PN camera within the 80 per cent EEF ellipse.

\subsection{X-ray optical identification parameters}

\noindent 
SDSS\_ID (20A): SDSS source identification number.

\noindent 
RA\_OPT (E): SDSS optical counterpart Right Ascension (J2000).

\noindent 
DEC\_OPT (E): SDSS optical counterpart Declination (J2000).

\subsection{SDSS photometry parameters} 

\noindent 
petro\_u (E): SDSS $u$-band petrosian magnitude of the optical
counterpart.

\noindent 
petro\_g (E): SDSS $g$-band petrosian magnitude of the optical
counterpart.

\noindent 
petro\_r (E): SDSS $r$-band petrosian magnitude of the optical
counterpart.

\noindent 
petro\_i (E): SDSS $i$-band petrosian magnitude of the optical
counterpart.

\noindent 
petro\_z (E): SDSS $z$-band petrosian magnitude of the optical
counterpart.

% Don't change these lines
\bsp    % typesetting comment
\label{lastpage}
\end{document}